\documentclass[letterpaper,journal]{IEEEtran}

\usepackage{amsmath,amsfonts}
\usepackage{array}
\usepackage{booktabs}
\usepackage{cite}
\usepackage{graphicx}
\usepackage{mathtools}
\usepackage{url}
\usepackage{tikz}
\usepackage{xcolor}
\usetikzlibrary{arrows.meta,calc,positioning}

\newcommand{\model}{\textsc{SubRoFormer}}
\newcommand{\smallmodel}{SubRoFormer-S}
\newcommand{\basemodel}{SubRoFormer-B}
\definecolor{todored}{RGB}{155,25,25}
\definecolor{todobg}{RGB}{255,246,224}

\newcolumntype{L}[1]{>{\raggedright\arraybackslash}p{#1}}

\begin{document}

\title{\resizebox{\textwidth}{!}{SubRoformer: Subband Time-Domain Representation for Music Source Separation}}

\author{Chunyat Wu, Chuang Ding, Paris Smaragdis, Qiuqiang Kong%
\thanks{Corresponding author: Qiuqiang Kong.}%
\thanks{Chunyat Wu and Qiuqiang Kong are with The Chinese University of
Hong Kong (e-mail: wuchunyat@link.cuhk.edu.hk; qqkong@ee.cuhk.edu.hk).}%
\thanks{Chuang Ding is with Moonstep AI (e-mail: dingchuang@ydyj.xyz).}%
\thanks{Paris Smaragdis is with Massachusetts Institute of Technology
(e-mail: paris@mit.edu).}}

\markboth{Draft for IEEE/ACM Transactions on Audio, Speech, and Language Processing}{Author Name(s): SubRoFormer for Music Source Separation}

\maketitle

\begin{abstract}
Music source separation extracts vocals and instruments from a mixed recording,
enabling remixing, content production, and music analysis. Existing systems
commonly operate on waveforms or full-band spectrograms. Waveform inputs contain
many samples even in short segments, whereas a single spectrogram resolution
imposes a shared time--frequency resolution trade-off across frequency ranges.
We propose SubRoFormer, a Transformer-based system comprising a subband module
and a residual RoFormer module. The subband module decomposes the mixture into
frequency-localized, lower-rate signals for separation. It then reconstructs
the target waveforms from the estimated subbands. The residual RoFormer module
models temporal and cross-band relationships on temporally compressed features.
It combines this context with fine-rate features through gated residual fusion
to estimate source-specific complex masks.
On the MUSDB18-HQ test set, training only on MUSDB18-HQ yields an average
signal-to-distortion ratio (SDR) of
10.63~dB across four stems using target-specific models, achieving
state-of-the-art performance among the compared systems without extra training
data. With 500 additional training songs, a single model jointly separating all
four stems achieves an average SDR of 11.59~dB, establishing state-of-the-art
performance among the compared single-model multi-stem systems trained with
extra data. Code and checkpoints are released at
\url{https://github.com/ArchiMickey/SubRoformer}.
\end{abstract}


\section{Introduction}
\IEEEPARstart{M}{usic} source separation (MSS) estimates the constituent
signals of a musical mixture, such as vocals, bass, drums, and other
instruments. The resulting stems support source-level rebalancing and
remixing, selective restoration or remastering, and stereo-to-surround
upmixing or spatial repositioning when original multitracks are
unavailable~\cite{cano2019mss,openunmix}. They also enable karaoke and can act
as preprocessing for automatic transcription, pitch tracking, chord analysis,
and other tasks in music information
retrieval~\cite{rafiioverview,huangmonaural}.

MSS remains a challenging problem because a music recording contains multiple
sources mixed into only one or two audio channels. Different instruments often
overlap in both time and frequency, making it difficult to determine which part
of the mixture belongs to each source. The problem becomes more challenging
because the sound of an instrument can vary substantially with pitch,
articulation, and recording conditions. Traditional methods rely on assumptions
such as harmonicity, source independence, or repetition, but these assumptions
do not always hold for diverse music recordings. Supervised neural networks can
instead learn source characteristics from data, but they require paired mixtures
and isolated stems, which are much less available than finished music
recordings~\cite{cano2019mss,rafiioverview,huangmonaural}. Effective MSS therefore
requires models that can capture both source characteristics and musical context
while preserving fine-grained acoustic details such as pitch, transients, and
phase.

The priors used for separation have evolved across several methodological
generations. In the 1990s, auditory scene analysis framed separation as
grouping acoustic components into coherent streams, while blind methods used
statistical independence and frequency-domain
unmixing~\cite{bregman1990,bell1995,smaragdis1998}. During the 2000s,
music-specific models imposed harmonic and spectral-continuity structure;
nonnegative matrix factorization~\cite{virtanenklapuri2002,virtanennmf}
represented a monaural spectrogram through spectral templates and temporally
sparse activations. The 2010s shifted toward supervised neural estimation of
time--frequency masks~\cite{huangmonaural,openunmix} and, later, multiscale
waveform synthesis~\cite{waveunet,demucs}.
In the 2020s, hybrid waveform--spectrogram systems and band-split recurrent or
Transformer models combined complementary representations with long-context
time--band modeling~\cite{hybriddemucs,htdemucs,bsrnn,bsroformer}. These
generations differ in assumptions, supervision, and output contract rather
than forming a sequence of directly comparable performance gains.

Across these approaches, a central design question is how to retain
spectral structure and local temporal detail while limiting the sequence
length processed by the separator. Waveform models receive high-rate
sample sequences and typically reduce their length through learned
encoders. Systems based on a full-band short-time Fourier transform
(STFT) instead expose frequency structure on a common time--frequency
grid. BSRNN \cite{bsrnn} and BS-RoFormer \cite{bsroformer} group frequency bins after this transform,
whereas channel-wise subband systems filter and downsample the waveform
before computing local spectra~\cite{bsrnn,bsroformer,cws,cwspresunet}.
We investigate the latter analysis strategy together with a separator
that estimates masks on a fine temporal grid but models longer-range
context on a shorter sequence.

Our design separates two forms of temporal reduction. First, subband
analysis lowers the sampling rate of each band-limited waveform while
retaining an explicit synthesis path. Second, the separator groups
neighboring local spectral frames into temporal patches for its main
Transformer. Here, the fine rate refers to the local spectral frame rate
at which masks are predicted, whereas the coarse rate refers to the
patch rate used for long-context modeling. A skip pathway retains
fine-rate features for mask prediction. Thus, waveform rate reduction
and feature-sequence reduction serve different purposes and are
controlled at different stages of the system.

Classical analysis filter banks provide the basis for this
frontend~\cite{multiratetutorial}. Each selected frequency band is
converted to a complex signal and shifted toward zero frequency by
demodulation. Its controlled bandwidth then permits processing at a
lower sampling rate. We call these complex baseband sequences the
subband time-domain representation; the separator operates on local
STFTs computed from them~\cite{stftanalysis}. After complex-mask
estimation, inverse local STFTs recover the estimated baseband signals,
and subband synthesis returns them to their original frequency ranges
and combines them into a waveform. Within the separator, fine-rate
processing and a skip pathway support mask estimation, while a
temporally patched Transformer models longer-range dependencies.
The frontend has an explicit synthesis path, but its finite-filter
implementation introduces reconstruction error, which we quantify
separately from separation performance.

To address these limitations, This work makes three contributions:
\begin{itemize}
    \item We formulate a fixed subband time-domain representation based
    on complex demodulation and rate reduction, with an explicit
    synthesis path and a quantitative assessment of reconstruction
    distortion.
    \item We develop a polyphase analysis--synthesis implementation that
    avoids computing full-rate samples discarded by decimation and
    avoids explicitly constructing zero-filled sequences during
    interpolation.
    \item We propose \model, which combines fine-rate feature processing
    and complex-mask prediction with a temporally patched axial
    RoFormer. A gated residual fusion module combines fine-rate features
    with the coarse-rate context representation.
    \item Under MUSDB18-HQ dataset, \model{} outperforms the compared systems without extra data.
\end{itemize}

On the MUSDB18-HQ test set, four separately trained, target-specific
\basemodel{} models achieve a four-stem mean signal-to-distortion ratio
(SDR) of 10.63~dB without extra training data. This is the highest
reported mean among the complete four-stem results in
Table~\ref{tab:baseline-comparison}. With 500 additional training songs,
a single \basemodel{} jointly predicts all four stems and reaches
11.59~dB, the highest reported mean among the compared single-model
multi-stem systems in Table~\ref{tab:baseline-comparison-extra}.
For our results, the four-stem mean averages the per-stem median SDRs
over the 50 test tracks. These comparisons summarize the reported
systems rather than isolate architectural effects under matched
training conditions.

Figure~\ref{fig:architecture} summarizes the complete analysis,
separation, and synthesis pipeline. Section~III describes the fixed
subband representation and its analysis--synthesis implementation,
whereas Section~IV presents the learned separator. Sections~V and VI
describe the experimental setup and results, respectively.

\section{Related Work}

Music source separation estimates individual source signals from their
mixture~\cite{cano2019mss}. Under an additive mixing model, let
\begin{equation}
    \mathbf{x}[n] = \sum_{j=1}^{J}\mathbf{s}_j[n],
    \qquad \mathbf{x}[n],\mathbf{s}_j[n]\in\mathbb{R}^{C \times L},
    \label{eq:mss-mixture}
\end{equation}
where $n$ indexes samples, $J$ is the number of sources, $L$ is the
number of time samples, and $C$ is the
number of audio channels. The objective is to estimate a target waveform
$\hat{\mathbf{s}}_j$ from $\mathbf{x}$, either with a source-specific model
or a model that jointly estimates multiple sources. We review waveform-based and hybrid systems first, followed by systems
that operate on explicit time--frequency representations. We then
focus on subband and nonuniform analysis, which are particularly
relevant to our frontend. These groups are not mutually exclusive:
a subband system may still perform separation on local spectrograms.
The organization distinguishes how the signal is represented from
how the network models dependencies within that representation.

Evaluation campaigns provide complementary context for MSS research:
SiSEC 2018 established a shared music-separation evaluation
setting~\cite{sisec2018}, MDX 2021 introduced a professionally created
hidden test set~\cite{mdx2021}, and SDX 2023 additionally examined
robustness to training-label errors and source bleeding~\cite{sdx2023}.

\subsection{Time-domain MSS Systems}
Waveform-only systems learn a mapping from mixture samples to source
samples without a fixed spectral transform~\cite{waveunet,demucs}:
\begin{equation}
    \hat{\mathbf{s}}_j = f_{\theta,j}(\mathbf{x}),
    \label{eq:mss-waveform}
\end{equation}
where $\theta$ denotes learned parameters and $j$ selects the target output.
Phase is implicit in the predicted waveform rather than estimated as a
separate spectral quantity. The central modeling challenge is to capture
long temporal dependencies at audio sampling rates. Wave-U-Net \cite{waveunet} and Demucs~\cite{demucs}
address this challenge through temporal downsampling and multiscale
encoder--decoder representations.

Wave-U-Net \cite{waveunet} repeatedly downsamples and upsamples temporal features, using
skip connections to combine information across resolutions. Its
context-aware prediction framework reduces boundary artifacts when
processing successive waveform segments~\cite{waveunet}. Demucs \cite{demucs} combines
strided convolutions with a bidirectional long short-term memory (LSTM)
module at the bottleneck, allowing temporal dependencies to be modeled on
a shorter latent sequence~\cite{demucs}. Both designs retain
high-resolution encoder features through skip connections rather than
performing all sequence modeling at the input sampling rate.

Speech-separation studies provide related strategies for latent-domain
masking and long-sequence processing, although their results are not
directly comparable to MSS benchmarks.
Conv-TasNet \cite{convtasnet} learns an analysis encoder and synthesis decoder around a temporal
convolutional masking network. Dual-path RNN~\cite{dprnn} instead
factorizes long-sequence processing into intra- and inter-chunk recurrent
operations, separating modeling within short segments from modeling across
segments.

Subsequent Demucs \cite{hybriddemucs, htdemucs} systems extend this waveform lineage with an explicit
spectral branch and are therefore hybrid rather than purely time-domain.
Hybrid Demucs couples waveform and complex-spectrogram branches through
a shared inner representation~\cite{hybriddemucs}. HT Demucs instead uses
a cross-domain Transformer to exchange information between the two
branches~\cite{htdemucs}. KUIELab-MDX-Net\cite{kuielabmdx} combines the separated outputs of
independent waveform and spectrogram streams~\cite{kuielabmdx}. The key
distinction is therefore not waveform output, which spectral systems also
produce, but whether separation uses learned temporal features alone or
combines them with an explicit spectral representation.

\subsection{Time--frequency-domain MSS Systems}
Time--frequency systems expose an explicit frequency axis by applying a
transform before separation. For a channel-wise short-time Fourier
transform (STFT), define $\mathbf{X}=\operatorname{STFT}(\mathbf{x})
\in\mathbb{C}^{C\times F\times T}$, where $F$ and $T$ denote frequency
bins and time frames, respectively~\cite{stftanalysis}. Complex-mask
systems estimate source coefficients and reconstruct waveforms as
\begin{equation}
    \hat{\mathbf{S}}_j = \mathbf{M}_j\odot\mathbf{X},
    \qquad
    \hat{\mathbf{s}}_j = \operatorname{iSTFT}(\hat{\mathbf{S}}_j),
    \label{eq:mss-spectral-mask}
\end{equation}
where $\mathbf{M}_j\in\mathbb{C}^{C\times F\times T}$ is a predicted
mask, $\hat{\mathbf{S}}_j$ contains the estimated source coefficients,
$\odot$ denotes element-wise multiplication, and $\operatorname{iSTFT}$
is the inverse STFT. Each complex mask coefficient scales the magnitude
and shifts the phase of the corresponding mixture coefficient.
Unlike a bounded real-valued mask, complex ideal ratio masks (cIRMs)~\cite{complexmask,bsrnn} are not restricted to retaining
a fraction of the mixture magnitude.

Within this representation, one line of work develops multiscale spectral
features. Spectrogram U-Nets \cite{singingunet} combine encoder and decoder features for
singing-voice separation, and Spleeter~\cite{spleeter} provides
pretrained U-Net-based separation models. MMDenseLSTM~\cite{mmdenselstm}
combines multiscale convolutional and recurrent processing,
while D3Net~\cite{d3net} uses multiple dilation factors to expand convolutional
receptive fields. Phase-aware ResUNet~\cite{phaseresunet} additionally estimates
magnitude and phase corrections rather than restricting reconstruction
to the mixture phase.

Source interaction and target conditioning form a complementary design
axis. X-UMX~\cite{xumx} bridges
source-specific networks and combines time- and frequency-domain training
objectives. LaSAFT~\cite{lasaft} instead uses source-conditioned frequency
transformations and gated feature modulation. These approaches
address source interaction and conditioning, complementing the time--band
factorization used by more recent band-split systems.

A further distinction concerns how frequency information is organized.
Open-Unmix~\cite{openunmix} uses a source-specific bidirectional LSTM to predict target
magnitudes from full-band features, followed by multichannel Wiener
filtering. BSRNN \cite{bsrnn} instead partitions the mixture STFT into
unequal-width bands chosen according to the target source. Band-specific
projections map these bands to equal-dimensional embeddings. Recurrent
blocks then alternate between time modeling within each band and
cross-band modeling within each frame, before band-specific heads predict
complex masks~\cite{bsrnn}. Thus, frequency grouping becomes explicit in
the separator rather than being absorbed into a full-band feature vector.

BS-RoFormer~\cite{bsroformer,roformer} retains band-specific embeddings but replaces recurrent
processing with Transformers that alternate attention along the time and
band axes, using rotary position embeddings.
Mel-RoFormer~\cite{melroformer} changes the nonoverlapping frequency partition to overlapping
mel-scale bands. SCNet~\cite{scnet} instead applies different
compression ratios across frequency regions, while DTTNet~\cite{dttnet}
combines time--frequency convolutions and frequency projections with
dual-path recurrent modeling. These methods differ in
frequency allocation and dependency modeling, but retain an STFT computed
from the full-band waveform as the analysis interface. This shared
interface distinguishes spectral band splitting from the waveform
subband analysis considered next.

\subsection{Subband and Nonuniform Representations}
Waveform subband analysis filters
the signal into frequency-localized sequences and may combine this
operation with decimation and synthesis~\cite{multiratetutorial}.
Subband decomposition does not itself require nonuniform bandwidths.
By contrast, the constant-Q transform uses narrower frequency bands
at low frequencies and wider bands at high
frequencies~\cite{browncqt}. Nonstationary Gabor frames provide a
framework for adaptive analysis and reconstruction~\cite{nsgabor},
including invertible constant-Q and sliced constant-Q
transforms~\cite{invertiblecqt}. 

Channel-wise subband input uses analysis--synthesis filtering and stacks subband spectra along the channel axis~\cite{cws}. CWS-PResUNet filters and downsamples waveform subbands before their STFTs and applies phase-aware separation to the resulting features~\cite{cwspresunet}. The original XUMX-SliCQ~\cite{xumxslicq} study replaces the STFT with a variable-resolution transform but reports lower separation scores than Open-Unmix; a subsequent system revisits this representation
for the Cadenza Challenge~\cite{xumxslicqv2}. MDCT \cite{jiang2024mdctcodec} has been employed to decompose audio signals into uniformly spaced frequency bands. However, these subband methods employ \textit{uniform frequency bands}, limiting their flexibility in adapting frequency resolution to different spectral regions.


\begin{figure*}[!t]
    \centering
    \includegraphics[width=\textwidth]{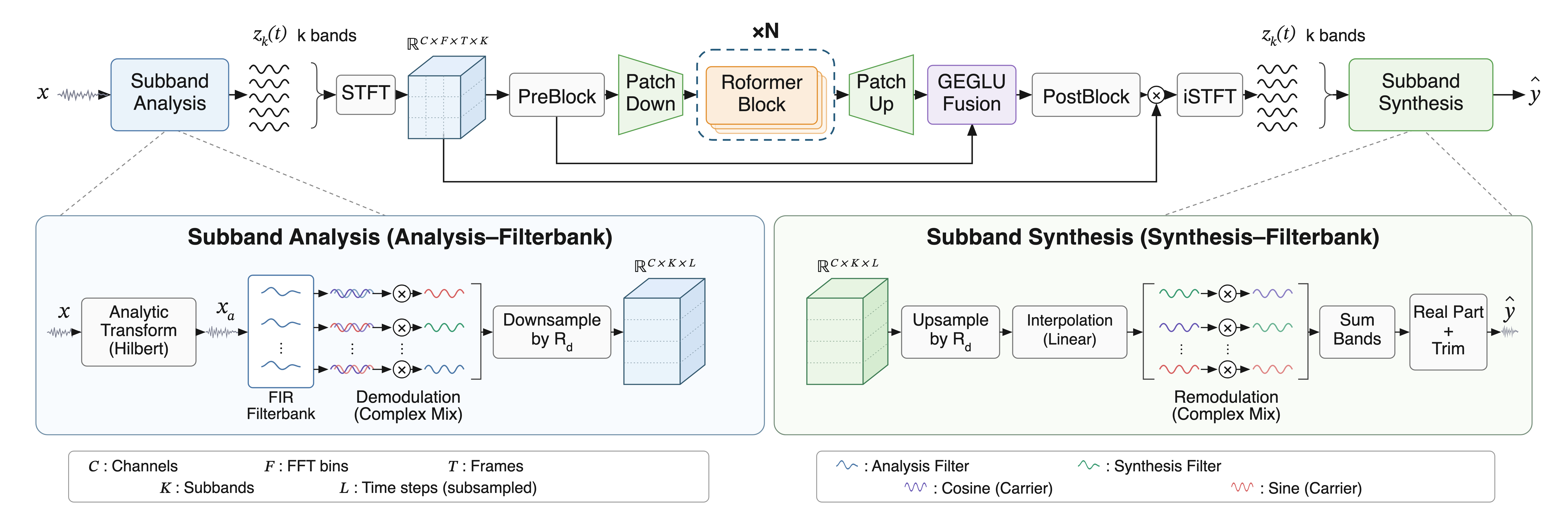}
    \caption{Overview of \model. The fixed front end converts the analytic
    mixture into demodulated, decimated subbands; local complex features are
    processed by fine-rate boundary blocks and a temporally patched axial
    RoFormer; the inverse path interpolates, remodulates, and sums the estimated
    subbands. $C$, $L$, and $K$ denote channels, time samples, and realized
    subbands, respectively.}
    \label{fig:architecture}
\end{figure*}

\section{Subband Time-Domain Representation}
Let $x[n]\in\mathbb{R}^{L}$ be a discrete-time waveform sampled at $F_s$, and let $K$ denote the number of subbands. We omit the channel dimension $C$ for simplicity, as the subband analysis is applied independently to each channel. The \textit{analysis transform} maps the input waveform $x[n]$ into $K$ \textit{complex-valued, low-rate baseband signals with arbitrarily chosen, non-uniform bandwidths}, represented by $\mathbf{z} \in \mathbb{C}^{K \times \frac{L}{D}}$, where $D$ denotes the decimation factor. The \textit{synthesis transform} reverses this rate conversion and combines the reconstructed bands. This representation retains within-band magnitude and phase while allowing each band to be processed below the full waveform rate. We describe the subband analysis and subband synthesis modules as below. 

\subsection{Subband Analysis}
The analysis module consists of four stages. To avoid aliasing and preserve the desired spectrum~\cite{multiratetutorial}, we first apply a band-pass filter to the input signal, convert the filtered signal into a complex analytic signal, translate its carrier frequency to zero, and then decimate it. These four operations transform a long, full-rate waveform sequence into a short, low-rate baseband sequence.




\subsubsection{Band-Pass Filtering}

\begin{figure}[t]
    \centering
    \includegraphics[width=\columnwidth]{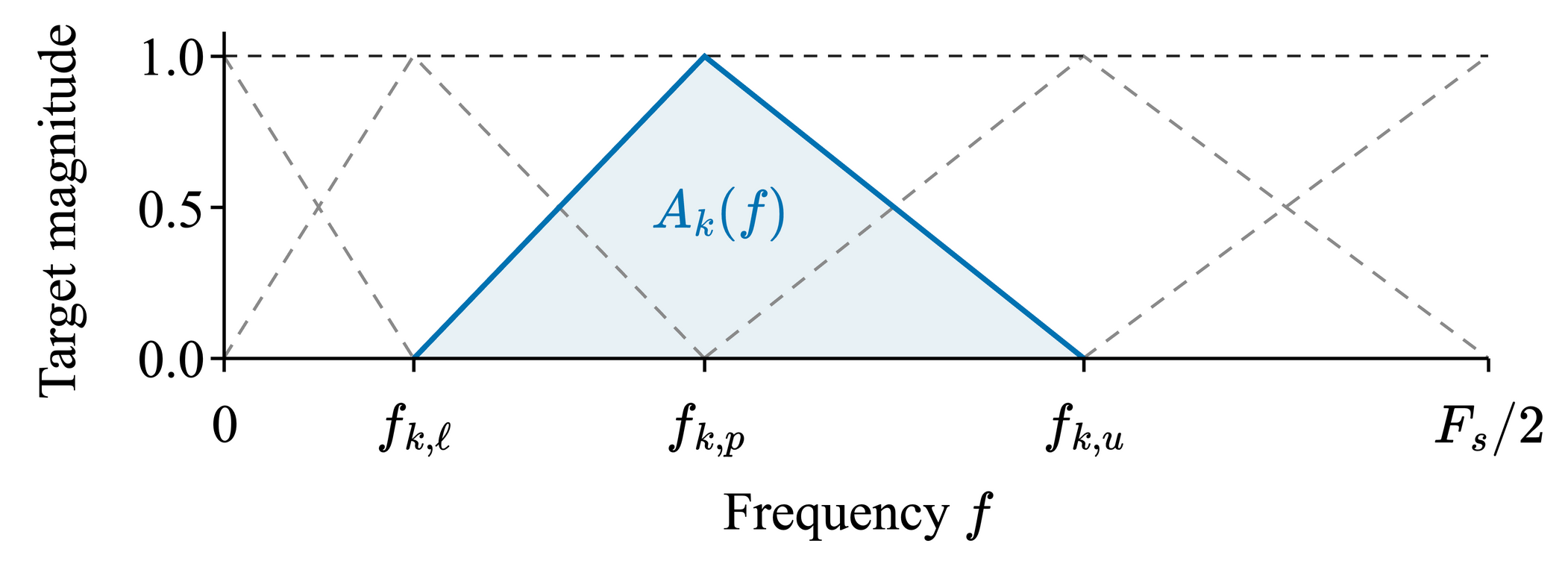}
    \caption{Schematic target magnitude responses of the triangular filter
    bank, with illustrative frequency spacing. The highlighted interior band
    $A_k(f)$ has lower edge $f_{k,\ell}$, peak $f_{k,p}$, and upper edge
    $f_{k,u}$. Neighboring responses overlap along their slopes; the first
    and last bands are one-sided. The dashed line shows their unit sum over
    $0\leq f\leq F_s/2$. These are ideal targets, rather than the realized
    finite-length FIR responses.}
    \label{fig:triangular-filterbank}
\end{figure}

The band-pass filters determine the frequency resolution of the analysis. Unlike the STFT, which has a fixed frequency resolution, a subband filter bank can provide \textit{arbitrary frequency resolutions} by changing the filter bandwidths. This is useful for analyzing low-frequency signals. The filters can have rectangular or triangular target magnitude responses, as long as they sum to unity over the nonnegative frequency range:
\begin{equation}
    \sum_{k=1}^{K} A_k(f) = 1, \qquad 0\leq f\leq F_s/2,
    \label{eq:subband-partition-unity}
\end{equation}
where $A_k(f)$ denotes the target magnitude response of band $k$, and $ F_{s} $ denotes sampling rate.
Here, we use a triangular filter bank. For an interior subband $k$, let $f_{k,\ell}$,
$f_{k,p}$, and $f_{k,u}$ denote its lower edge, response peak, and upper edge.
Its target magnitude response is
\begin{equation}
    A_k(f) =
    \begin{cases}
        \dfrac{f-f_{k,\ell}}{f_{k,p}-f_{k,\ell}},
            & f_{k,\ell}\leq f\leq f_{k,p},\\[6pt]
        \dfrac{f_{k,u}-f}{f_{k,u}-f_{k,p}},
            & f_{k,p}<f\leq f_{k,u},\\[6pt]
        0, & \text{otherwise},
    \end{cases}
    \label{eq:subband-triangular-response}
\end{equation}
over the nonnegative-frequency interval. An advantage of using the triangular responses is that adjacent bands overlap along their slopes, avoiding the abrupt boundaries of disjoint rectangular bands. The first and last bands use the corresponding one-sided low-pass and high-pass responses, so that the filter bank covers the complete spectrum.
Figure~\ref{fig:triangular-filterbank} illustrates these target responses and
their overlap.

We implement each of the $K$ band-pass filters using a finite impulse response (FIR) filter. A Hamming window is used to control the sidelobes introduced by finite-length truncation~\cite{oppenheimdsp}. We implement the FIR filters using \textit{firwin2} from SciPy~\cite{scipy2020}. The frequency response of the $k$-th bandpass filter is denoted by $A_k(f)$. All filter coefficients are fixed after construction rather than learned jointly with the separator. For each subband $k$, we apply the corresponding band-pass filter to $x[n]$ to obtain the band-pass filtered signal $x_k[n]$. 

\begin{figure}[t]
    \centering
    \includegraphics[width=\columnwidth]{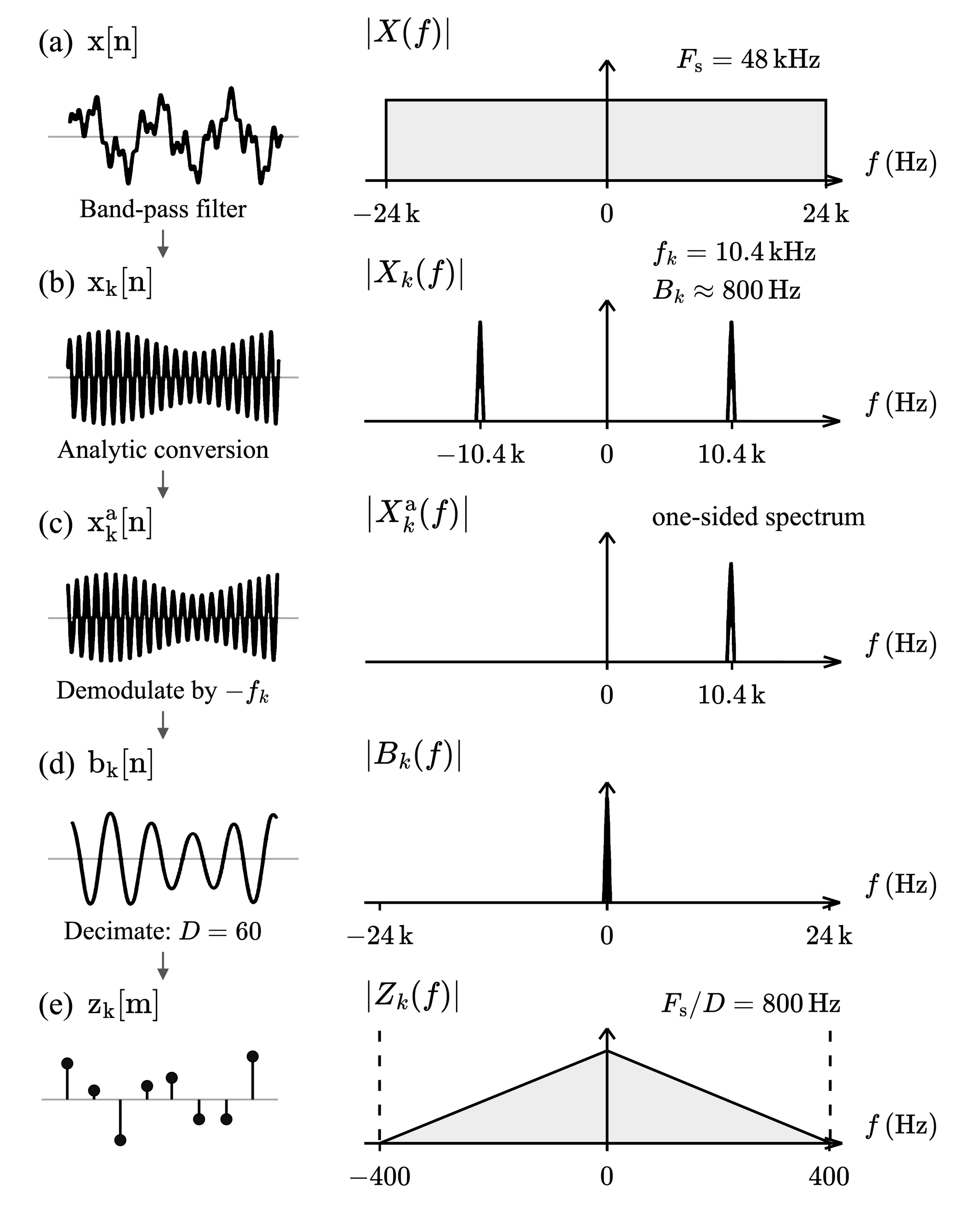}
    \caption{Per-band analysis and rate conversion: (a) the input $x[n]$;
    (b) the band-pass filtered signal $x_k[n]$; (c) the analytic signal
    $x_k^{\mathrm{a}}[n]$; (d) the complex baseband signal $b_k[n]$; and
    (e) the decimated sequence $z_k[m]$. Uppercase symbols denote the
    corresponding spectra. The example uses $F_s=48$~kHz, $f_k=10.4$~kHz,
    approximately 800-Hz band support, and $D=60$, giving a reduced sampling
    rate of $F_s/D=800$~Hz.}
    \label{fig:subband-rate-conversion}
\end{figure}

Figure~\ref{fig:subband-rate-conversion}(a) shows the input signal $x[n]$ and its magnitude spectrum. Figure~\ref{fig:subband-rate-conversion}(b) shows the band-pass filtered signal $x_k[n]$ for one band and its magnitude spectrum. The signal retains the band around the 10.4-kHz carrier, with a bandwidth of approximately 800 Hz. This bandwidth restriction makes the subsequent rate reduction possible.


\subsubsection{Analytic Conversion}
The frequency response of $x_k[n]$ is symmetric, so its negative-frequency component is redundant. We therefore convert the real-valued band signal to its analytic counterpart~\cite{oppenheimdsp}:
\begin{equation}
    x_k^{\mathrm{a}}[n]
    = x_k[n] + j\mathcal{H}\{x_k[n]\},
    \label{eq:subband-analytic}
\end{equation}
where $\mathcal{H}$ denotes the Hilbert transform. The analytic signal suppresses the negative-frequency components and retains the positive-frequency components as shown in Figure~\ref{fig:subband-rate-conversion}(c).


\subsubsection{Baseband Demodulation}
Let $f_k$ be the center frequency of subband $k$. By using the Fourier
frequency-shift property~\cite{oppenheimdsp}
\begin{equation}
    x[n]e^{j\omega_0 n}
\;\overset{\mathcal{F}}{\longleftrightarrow}\;
X(\omega-\omega_0)
    \label{eq:frequency-shift}
\end{equation}
where $X(\omega)$ is the discrete-time Fourier transform of $x[n]$ and
$\omega$ is angular frequency in radians per sample. Complex demodulation
with $\omega_0=-2\pi f_k/F_s$ translates the carrier to zero frequency:
\begin{equation}
    b_k[n]
    = x_k^{\mathrm{a}}[n]e^{-j2\pi f_k n/F_s},
    \label{eq:subband-demodulation}
\end{equation}
where $b_k[n]$ denotes the complex baseband signal. Figure~\ref{fig:subband-rate-conversion}(d) shows $b_k[n]$ and its magnitude spectrum. The narrow bandwidth allows the signal to be decimated in the next stage.

\subsubsection{Decimation}
The demodulated signal is retained every $D$ samples,
\begin{equation}
    z_k[m] = b_k[mD].
    \label{eq:subband-decimation}
\end{equation}
If the total baseband support has width $B_k < F_s/D$, the decimation is alias-free according to the Nyquist sampling theorem~\cite{multiratetutorial}, and the signal can be perfectly reconstructed. Since the same decimation factor $D$ is used for all subbands, it must be chosen based on the widest subband. In practice, finite filter transition bands can introduce residual aliasing. We therefore allow a small tolerance in the bandwidth, such as using 780 Hz for a target bandwidth of 800 Hz. Figure~\ref{fig:subband-rate-conversion}(e) shows the decimated signal $z_k[m]$ and its magnitude spectrum. 

Taken together, stacking the subband signals $z_{k}[m]$ across $C$ channels and $K$ bands gives a subband representation $\mathbf{z}\in\mathbb{C}^{C\times K\times \frac{L}{D}}$, assuming $L$ is divisible by $D$.


\subsection{Subband Synthesis}

The subband synthesis module mirrors the analysis module. It takes $\mathbf{z}\in\mathbb{C}^{C\times K\times \frac{L}{D}}$ as input and reconstructs the original signal $x\in\mathbb{R}^{C\times L}$. The subband synthesis module consists of four operations that are symmetric to those in the subband analysis module: interpolation, shifting back to the original frequency, conversion to a real-valued band contribution, and summation across all bands.

\subsubsection{Interpolation}
For each subband signal $z_{k}[m]\in \mathbb{R}^{\frac{L}{D}}$, synthesis first forms the factor-$D$ upsampled sequence $z_k^{\mathrm{up}}[n] \in \mathbb{R}^{L}$ by inserting $D-1$ zeros between adjacent samples.
\begin{equation}
    z_k^{\mathrm{up}}[n]
    =
    \begin{cases}
        z_k[n/D], & n\mod D = 0,\\
        0, & \text{otherwise}.
    \end{cases}
\end{equation}
Then, $z_k^{\mathrm{up}}[n]$ is passed through an interpolation filter
$g[n] = D\operatorname{sinc}\left(\frac{n}{D}\right)$, which is truncated by a Hamming window. The interpolated signal $u_k[n] \in \mathbb{R}^{L}$ is given by
\begin{equation}
    u_k[n]
    = \left(z_k^{\mathrm{up}}*g\right)[n].
    \label{eq:subband-interpolation}
\end{equation}
as shown from Fig.~\ref{fig:subband-rate-conversion}(e) to Fig.~\ref{fig:subband-rate-conversion}(d).

\subsubsection{Remodulation}

For each interpolated baseband signal $u_k[n]\in\mathbb{R}^{L}$, synthesis shifts its spectrum back to the original frequency location using the frequency-shift property in~(\ref{eq:frequency-shift}) with $\omega_0=2\pi f_k/F_s$. The remodulated analytic band signal $v_k[n]\in\mathbb{C}^{L}$ is given by
\begin{equation}
    v_k[n] = u_k[n]e^{j2\pi f_k n/F_s}.
    \label{eq:subband-remodulation}
\end{equation}
As shown from Fig.~\ref{fig:subband-rate-conversion}(d) to Fig.~\ref{fig:subband-rate-conversion}(c), the interpolated baseband signal is shifted back to its original spectral location.

\begin{table*}[t]
    \centering
    \caption{Transform reconstruction quality on the MUSDB18-HQ test set.
    SDR is delay-aligned and summarized by the median over 50 songs per
    signal; LSD and NME are medians over 250 song--signal pairs.
    Rate counts real scalar coefficients per sample per channel, including
    padding. Higher SDR and lower LSD and NME indicate better fidelity.}
    \label{tab:transform-reconstruction}
    \small
    \setlength{\tabcolsep}{4pt}
    \begin{tabular}{@{}lrrrrrrrr@{}}
        \toprule
        & & \multicolumn{5}{c}{Reconstruction SDR (dB) $\uparrow$}
        & & \\
        \cmidrule(lr){3-7}
        Transform & Rate & Mixture & Vocals & Bass & Drums & Other
        & LSD (dB) $\downarrow$ & NME $\downarrow$ \\
        \midrule
        STFT~\cite{stftanalysis} & 4.271 & 138.58 & 138.47 & 138.75 & 138.64 & 138.63
        & 0.0000 & 0.0000 \\
        \textbf{SubRoFormer mel--linear} & 4.267 & 28.04 & 26.80 & 47.78 & 26.58 & 29.03
        & 3.3375 & 0.0264 \\
        SubRoFormer ERB--linear & 4.267 & 28.04 & 26.80 & 48.26 & 26.56 & 29.11
        & 3.3318 & 0.0265 \\
        Auditory ERB~\cite{erbscale} & 0.125 & 0.54 & 0.05 & 1.00 & 0.88 & 0.15
        & 113.0755 & 0.9353 \\
        PQMF~\cite{multiratetutorial} & 1.000 & 11.30 & 10.25 & 13.41 & 12.03 & 10.38
        & 3.5195 & 0.2770 \\
        MDCT~\cite{mdct} & 1.000 & 132.03 & 132.16 & 131.77 & 131.98 & 132.15
        & 0.0001 & 0.0000 \\
        Haar wavelet packet~\cite{waveletpackets} & 1.000 & 131.57 & 131.57 & 131.58 & 131.55 & 131.59
        & 0.0000 & 0.0000 \\
        \bottomrule
    \end{tabular}
\end{table*}

\subsubsection{Real-Valued Projection}
For each analytic band signal $v_k[n]\in\mathbb{C}^{L}$, synthesis takes its real part to obtain the real-valued band contribution $y_k[n]\in\mathbb{R}^{L}$:
\begin{equation}
    y_k[n] = \operatorname{Re}\{v_k[n]\}.
\end{equation}
As shown from Fig.~\ref{fig:subband-rate-conversion}(c) to Fig.~\ref{fig:subband-rate-conversion}(b), the analytic band signal is projected back to a real-valued signal.

\subsubsection{Band Summation}
For all $K$ subbands, synthesis sums their real-valued band contributions to reconstruct the waveform $y[n]\in\mathbb{R}^{L}$:
\begin{equation}
    y[n] = \sum_{k=1}^{K} y_k[n].
    \label{eq:subband-summation}
\end{equation}
As shown from Fig.~\ref{fig:subband-rate-conversion}(b) to Fig.~\ref{fig:subband-rate-conversion}(a), the band contributions are summed to reconstruct the waveform.




\subsection{Degradation Analysis}
Ideally, the subband analysis and synthesis modules will lead to perfect reconstruction. However, due to bandpass filter aliasing, and ringing in the edges effect, there are distortions. We quantify distortion in the fixed analysis--synthesis path using the stereo mixture and four reference stems from all 50 MUSDB18-HQ test songs~\cite{musdbhq}. Audio is 48~kHz and compared with this resampled reference, excluding resampling loss. Table~\ref{tab:transform-reconstruction} reports delay-aligned reconstruction SDR, log-spectral distance (LSD), normalized magnitude error (NME), and coefficient rate. These metrics evaluate reconstruction quality from different aspects. The reconstruction SDR measures the fidelity of the fixed analysis--synthesis transform.

The mel--linear and ERB--linear banks, based on the mel and equivalent rectangular bandwidth (ERB) scales~\cite{melscale,erbscale}, show similar distortion patterns, with bass reconstructed more accurately than the other stems. Their coefficient rates are almost identical to that of the tested STFT, but their reconstruction SDRs are much lower. In contrast, the STFT, MDCT, and Haar controls achieve near-perfect reconstruction.

\subsection{Fast Polyphase Implementation}


A direct implementation of the subband analysis and synthesis modules requires \(O(KNL)\) computational complexity, where \(K\) is the number of subbands, \(N\) is the filter length, and \(L\) is the input sequence length. This can become computationally expensive when using long FIR filters, which are needed to achieve the desired frequency selectivity. To improve efficiency, we propose an efficient polyphase implementation for both the subband analysis and synthesis modules.

As a first step, we implement all convolution operations using FFT-based convolution~\cite{multiratetutorial}. Since \(L \gg N\) in our setting, this reduces the complexity of each convolution from \(O(NL)\) to \(O(L\log L)\), and the overall complexity from \(O(KNL)\) to \(O(KL\log L)\). In the analysis module, the filtered signals are downsampled by a factor of \(D\). We therefore jointly implement band-pass filtering and decimation using polyphase convolution, avoiding computation of samples that are discarded after downsampling. This reduces the computational complexity and intermediate memory usage to \(O\left(K\frac{L}{D}\log\frac{L}{D}\right)\).

Similarly, in the subband synthesis module, the low-rate subband signals are first interpolated by a factor of \(D\) before filtering and remodulation. We exploit the polyphase structure to jointly implement interpolation and filtering, avoiding computation of intermediate samples that are not required. This reduces the computational complexity and intermediate memory usage to \(O\left(K\frac{L}{D}\log\frac{L}{D}\right)\).




\section{Residual SubRoFormer}
We perform source separation on the subband time-domain representation using
the proposed residual \model{}. The design couples a fine-rate feature path
with a patched coarse-rate separator through residual fusion, following the
residual-learning principle of adding learned updates to an existing feature
stream~\cite{resnet}. The overall architecture is illustrated in Fig.~\ref{fig:architecture}.

\subsection{Residual Separator Architecture}
SubRoformer follows the axial time--band modeling principle of
BS-RoFormer~\cite{bsroformer} and comprises four components.
\begin{itemize}
    \item \textbf{Preprocessing Block:} refines fine-rate local features and
    retains a skip representation.
    \item \textbf{Time--Frequency RoFormer:} alternates time- and cross-band
    attention on temporally patched features.
    \item \textbf{GEGLU Fusion Module:} combines coarse- and fine-rate features
    through gated residual fusion.
    \item \textbf{Postprocessing Block:} refines the fused features and predicts
    target-specific complex masks.
\end{itemize}

\subsubsection{Preprocessing Block}
Let $\mathbf{z} \in \mathbb{C}^{C \times K \times (L/D)}$ denote the subband representation described in Section~III. We apply the STFT independently to each subband along the downsampled time axis, yielding $\mathbf{X} \in \mathbb{C}^{C \times K \times T \times F}$, where $T$ and $F$ denote the numbers of local time frames and frequency bins, respectively. The complex representation $\mathbf{X}$ is converted to a real-valued representation by concatenating its real and imaginary parts. Then, the channel and frequency dimensions are flattened into a single dimension. The preprocessing and postprocessing blocks each contain a ConvNeXt-V2-style temporal mixer~\cite{convnextv2} with RMSNorm~\cite{rmsnorm} and SwiGLU gating. It is then processed by a Transformer layer along the temporal axis to capture temporal dependencies within each subband, followed by another Transformer layer along the subband axis to model interactions across subbands at each time frame. The resulting representation captures both temporal and cross-band dependencies at fine resolution. The output is denoted as $\mathbf{X}' \in \mathbb{R}^{K \times T \times D}$, where $D$ is the latent dimension.



A learned patchification module groups neighboring temporal frames of $\mathbf{X}'$ to reduce the sequence length while preserving the subband axis, yielding $\mathbf{H}^{(0)} \in \mathbb{R}^{K \times T' \times D'}$, where $T'$ and $D'$ denote the sequence length after patchification and the latent dimension, respectively. We apply patchification only along the temporal axis because neighboring temporal frames contain redundant information. In contrast, we do not apply patchification along the subband axis because further compression of subband information may lead to the loss of spectral details for source separation. 


\subsubsection{Time--Frequency RoFormer}

The main separator takes the temporally patched representation $\mathbf{H}^{(0)}$ as input and consists of stacked axial RoFormer blocks. Each block contains a subband-axis Transformer $\mathcal{B}_{\ell}$ and a temporal-axis Transformer $\mathcal{T}_{\ell}$, applied sequentially as
\begin{equation}
    \mathbf{H}^{(\ell)} = \mathcal{T}_{\ell}\left(\mathcal{B}_{\ell}\left(\mathbf{H}^{(\ell-1)}\right)\right),
    \label{eq:separator-axial}
\end{equation}
where $\mathcal{B}_{\ell}$ and $\mathcal{T}_{\ell}$ denote the subband-axis and temporal-axis Transformer blocks at layer $\ell$, respectively. The axial RoFormer body follows the BS-RoFormer design~\cite{bsroformer} with SwiGLU feed-forward networks~\cite{swiglu}. The subband-axis Transformer models interactions across subbands at each time frame. The temporal-axis Transformer captures long-range temporal dependencies within each subband. Both Transformers use independent parameters and rotary positional embeddings (RoPE)~\cite{roformer} to encode positional information along their respective axes. Each Transformer block includes self-attention, normalization, residual connections, and a feed-forward network. This axial design~\cite{axialattention} captures temporal and cross-band dependencies without flattening the entire time–subband grid. Operating on the temporally patched representation further reduces the computational cost of the main separator.


\subsubsection{GEGLU Fusion Module}
We apply a learned unpatchification module to map the coarse representation $\mathbf{H}^{(L)} \in \mathbb{R}^{K \times T' \times D'}$ back to the fine temporal resolution, yielding $\mathbf{Y}' \in \mathbb{R}^{K \times T \times D}$. The upsampled representation $\mathbf{Y}'$ is then combined with the preprocessing skip representation $\mathbf{X}'$ by concatenating them along the latent dimension, yielding $\mathbf{Y} = \operatorname{Concat}(\mathbf{X}', \mathbf{Y}')$. The concatenated representation is processed by a GEGLU fusion module~\cite{swiglu} to adaptively combine information from both representations:
\begin{equation}
\mathbf{Y}_{\mathrm{fuse}} = \mathbf{Y}' + \mathbf{W}_o\left(\operatorname{GELU}(\mathbf{W}_g\mathbf{Y}) \odot \mathbf{W}_v\mathbf{Y}\right),
\end{equation}
where $\mathbf{W}_g$, $\mathbf{W}_v$, and $\mathbf{W}_o$ are learnable projections, and $\odot$ denotes elementwise multiplication. This fusion combines long-range dependencies captured by the main separator with fine-resolution information preserved by the skip connection.

\subsubsection{Postprocessing Block}

The fused features are processed by a postprocessing block consisting of one temporal Transformer layer followed by one subband Transformer layer, symmetric to the preprocessing block. We denote this output as $ \mathbf{U} \in \mathbb{R}^{K \times T \times D} $. The output $\mathbf{U}$ is projected to a tensor of shape $K \times T \times 2CF$, then reshaped into channel, frequency, and real and imaginary components, formed a complex-valued mask $\mathbf{M} \in \mathbb{C}^{C \times K \times T \times F}$. The mask is applied to the mixture spectra $\mathbf{X}$ through elementwise multiplication:
\begin{equation}
    \hat{\mathbf{Z}} = \mathbf{M} \odot \mathbf{X},
\end{equation}
\noindent where $\odot$ denotes elementwise multiplication. We then apply the inverse STFT (iSTFT) to $\hat{\mathbf{Z}}$ to reconstruct the subband signals $\hat{\mathbf{z}} \in \mathbb{R}^{C \times K \times \frac{L}{D}}$. Finally, the subband synthesis described in Section~III-B is applied to $\hat{\mathbf{z}}$ to obtain the reconstructed waveform $\hat{\mathbf{s}} \in \mathbb{R}^{C \times L}$.

\subsection{Single-Stem and Multi-Stem Separation}
\label{sec:stem-configurations}
Section~IV-A describes the \textit{Single-stem separation}, which uses an individual model to separate each stem. Separating $J$ stems therefore requires $J$ different models and $J$ forward passes. \textit{Multi-stem separation} uses a shared backbone with separate output heads for different stems. Its advantage is that multiple stems can be separated in a single forward pass with fewer model parameters. Our proposed method supports both single-stem and multi-stem separation.

For multi-stem separation, we adopt the same backbone architecture but use separate output heads to predict multiple stems simultaneously. Specifically, the shared backbone extracts a latent representation $\mathbf{U} \in \mathbb{R}^{K \times T \times D}$, which contains information for separating different stems. To separate $J$ stems, $\mathbf{U}$ is projected to a tensor of shape $K \times T \times 2JCF$, which is then reshaped and converted into $J$ complex-valued masks $\mathbf{M}_{j} \in \mathbb{C}^{C \times K \times T \times F}$, where $j=1,\ldots,J$. Each mask corresponds to a specific stem and is applied to the mixture spectra $\mathbf{X}$ through elementwise multiplication:
\begin{equation}
    \hat{\mathbf{Z}}_{j} = \mathbf{M}_{j} \odot \mathbf{X}.
\end{equation}
The resulting $\hat{\mathbf{Z}}_{j}$ represents the estimated complex spectra of the $j$-th stem. We then apply the iSTFT to recover its subband signals, followed by the subband synthesis described in Section~III-B to reconstruct the waveform $\hat{\mathbf{s}}_{j} \in \mathbb{R}^{C \times L}$. In this way, the multi-stem system separates all $J$ stems in a single forward pass through the shared backbone, avoiding the need to run an individual model for each stem.

\subsection{Training Objective}
We train on 2-s stereo crops (96,000 samples per channel at 48~kHz).
Multi-resolution spectral objectives have been used for waveform
generation~\cite{parallelwavegan}; here, we combine waveform L1 error with
multi-resolution complex-STFT error:
\begin{equation}
    \mathcal{L} = \operatorname{mean}|\hat{y}-y|
    + \sum_{w\in\mathcal{W}}
    \operatorname{mean}|\mathcal{S}_{w}(\hat{y})-\mathcal{S}_{w}(y)|,
    \label{eq:training-loss}
\end{equation}
where $y$ and $\hat{y}$ denote the reference and estimated waveforms,
$\mathcal{W}=\{4096,2048,1024,512,256\}$, and $\mathcal{S}_{w}$ is an
unnormalized STFT with a Hann window of $w$ samples, a hop of 147 samples,
and an FFT size of $\max(w,2048)$. For complex residuals, $|\cdot|$ denotes
the complex modulus, not the sum of absolute real and imaginary components.
The waveform mean includes batch, stem, channel, and sample dimensions; each
spectral mean includes batch, stem, channel, frequency, and frame dimensions.
The five resolution-specific means are summed, not averaged, and the waveform
term and each resolution-specific term have unit weight. If
$\ell(\hat{\mathbf{s}}_j,\mathbf{s}_j)$ denotes this objective evaluated
for one stem, the two training configurations can be written as
\begin{equation}
    \begin{aligned}
        \mathcal{L}_{\mathrm{single},j}
            &= \ell(\hat{\mathbf{s}}_j,\mathbf{s}_j),\\
        \mathcal{L}_{\mathrm{multi}}
            &= \frac{1}{J}\sum_{j=1}^{J}
                \ell(\hat{\mathbf{s}}_j,\mathbf{s}_j).
    \end{aligned}
    \label{eq:stem-training-loss}
\end{equation}
The multi-stem expression is equivalent to including the stem dimension in
each mean in~\eqref{eq:training-loss}, giving every stem equal weight.

\section{Experiments}
\label{sec:experiments}

\subsection{Datasets}
Our main comparisons use the 100 training songs from MUSDB18-HQ, a 44.1-kHz
uncompressed version of the original MUSDB18 dataset~\cite{musdb,musdbhq}.
Each song contains four stereo stems: vocals, bass, drums, and other. All
four target-specific configurations use the augmentation policy from ZFTurbo's
Music-Source-Separation-Training~\cite{msst}. We retain these settings unchanged
to hold the augmentation recipe constant and maintain a fair comparison with
systems trained using the same recipe. Training mixtures are constructed by
independently sampling a track and temporal segment for each stem, applying the
augmentations below, and summing the resulting stems.
Each stem is independently scaled by a
gain sampled uniformly from $[0.5,1.5]$ and undergoes stereo-channel exchange,
time reversal, and polarity inversion with probabilities 0.5, 0.1, and 0.5,
respectively. Same-stem mixup starts from one source and independently adds up
to two further sources with probabilities 0.2 and 0.02; the components are
independently scaled in $[0.5,1.5]$ and averaged. Table~\ref{tab:vocals-ablations} gives the remaining
stem-dependent augmentations.


\subsection{Evaluation Metrics}
We use signal-to-distortion ratio (SDR) implemented by
\texttt{museval}~\cite{sisec2018,bsseval} as the primary separation metric.
The SDR is calculated by:
\begin{equation}
    \text{SDR} = 10 \log_{10} \frac{\|y\|^2}{\|\hat{y} - y\|^2},
\end{equation}
where $y$ is the reference signal and $\hat{y}$ is the estimated signal. The higher the SDR, the better the separation performance. The reported score for a stem is the median of these per-track values over all 50 MUSDB18-HQ test tracks. The four-stem mean is the arithmetic mean of the reported vocals, bass, drums, and other SDR values.

\subsection{Implementation Details}
\subsubsection{Subband Frontend}
All audio is resampled to 48~kHz for the proposed models, which operate on
stereo signals and use a fixed mel-to-linear triangular analysis bank.
The low-frequency bands are spaced on a mel scale~\cite{melscale}, whereas the
high-frequency
bands are spaced on a linear scale with 390-Hz intervals, yielding 128
overlapping subbands. After demodulation, each subband is
decimated by a factor of 60 to an 800-Hz complex baseband signal. We then apply
a two-sided 32-point Hann-windowed local STFT with a hop of eight subband
samples, producing a 100-frame/s fine temporal grid. The temporal patch factor
of four reduces the grid presented to the RoFormer body to 25 frames/s.

The 64-band ablation uses
790-Hz linear spacing in the upper-frequency region, yielding 64 realized
subbands. Its decimation factor is reduced to 30, giving a 1600-Hz subband
sample rate. The local STFT retains the 32-point Hann window but uses a hop of
16 subband samples, producing a 100-frame/s fine temporal grid. The temporal
patch factor remains four, yielding a 25-frame/s RoFormer-body grid and
matching both temporal rates of the 128-band reference.
All other separator settings are inherited from the 128-band \smallmodel{}
configuration.

\subsubsection{Separator Configuration}
The axial RoFormer body comprises six time--band layers with a model dimension
of 384 in \smallmodel{} and 12 layers with a model dimension of 576 in
\basemodel{}.

\subsubsection{Optimization and Training Budgets}
We optimize all proposed models with AdamW~\cite{adamw} using a peak learning
rate of $3\times10^{-4}$, $\beta_1=0.9$, $\beta_2=0.95$, and zero weight
decay. The learning rate is linearly warmed up for 1,000 optimizer updates,
then follows a cosine schedule~\cite{sgdr} without restarts, with a terminal
learning rate of zero.
Training uses bfloat16 mixed precision, no gradient accumulation, and a
maximum gradient norm of 1.0. An exponential moving average (EMA) of the
parameters is updated after each optimizer update with decay 0.999.

The \smallmodel\ runs and all ablations are trained on one NVIDIA GeForce
RTX 4090 with a batch size of four.
The MUSDB18-HQ-only \basemodel\ runs use six NVIDIA GeForce RTX 5090 GPUs
with a per-GPU batch size of three, giving a global batch size of 18.
For MUSDB18-HQ-only training, both single-stem and four-stem models are trained
for one million optimizer updates.
All reported results for the proposed models use the final EMA weights at the
end of each training run, rather than validation-selected checkpoints.
Full-track inference uses
2-s segments with 75\% overlap and Hamming-windowed overlap-add, evaluated
with bfloat16 precision.

\subsection{Baselines}
We compare our models with leading source separation systems including,
time-domain systems: Demucs V2 HQ~\cite{demucs}; time-frequency-domain systems:
BSRNN~\cite{bsrnn}, TFC--TDF~\cite{tfctdfv3}, SCNet~\cite{scnet},
BS-RoFormer~\cite{bsroformer}, and Mel-RoFormer~\cite{melroformer}; and hybrid
systems: HT Demucs~\cite{htdemucs} and Hybrid Demucs~\cite{hybriddemucs}.

Prior work uses different model configurations for four-stem evaluation:
multiple target-specific models separate individual stems, whereas a single
multi-stem model estimates all four stems in one forward pass. We evaluate
both target-specific and multi-stem versions of \model{} to support fairer
comparisons with baselines using either configuration.

Tables~\ref{tab:baseline-comparison}
and~\ref{tab:baseline-comparison-extra} include a multi-stem \basemodel{}
that jointly predicts vocals, bass, drums, and other with a single 97.23M
network ($N=1$). BS-RoFormer L12 instead uses three 93.4M target-specific
networks ($N=3$), or 280.2M parameters in total, to cover the same four
stems: it estimates vocals, bass, and drums and forms other as the mixture
residual~\cite{bsroformer}. The four-stem \basemodel{} therefore uses
roughly one third of the deployed parameters of that BS-RoFormer
configuration. This is a parameter-efficiency comparison of output contract
and model count, not a matched extra-data or checkpoint-generation study.

\begin{table}[t]
\caption{MUSDB18-HQ metrics for systems trained without extra data.
Metrics for research baselines are taken from their original papers.}
\label{tab:baseline-comparison}
\centering
\scriptsize
\setlength{\tabcolsep}{1.5pt}
\renewcommand{\arraystretch}{1.12}
\resizebox{\columnwidth}{!}{%
\begin{tabular}{@{}lccccccc@{}}
\toprule
Model & Params. & $N$ & Voc. $\uparrow$ & Bass $\uparrow$ & Dr. $\uparrow$ &
Oth. $\uparrow$ & Avg. $\uparrow$ \\
\midrule
\multicolumn{8}{c}{\textit{Single-stem models}} \\
\midrule
BSRNN~\cite{bsrnn} & 37.6M & 4 & 10.01 & 7.22 & 9.01 & 6.70 & 8.24 \\
BS-RoFormer~\cite{bsroformer,melroformer} & 82.8M & 4 & 11.02 & \textbf{11.58} & 9.66 & 7.80 & \underline{10.02} \\
Mel-RoFormer~\cite{melroformer} & 94.8M & 4 & \underline{11.60} & -- & 9.34 & 7.93 & -- \\
\textbf{\smallmodel{}} (Ours) & 22.52M & 4 & 11.34 & 9.18 & \textbf{10.60} & \underline{8.16} & 9.82 \\
\textbf{\basemodel{}} (Ours) & 97.17M & 4 & \textbf{12.63} & \underline{10.20} & \underline{10.40} & \textbf{9.27} & \textbf{10.63} \\
\midrule
\multicolumn{8}{c}{\textit{Multi-stem models}} \\
\midrule
HT Demucs~\cite{htdemucs} & 41.4M & 1 & 7.93 & 8.48 & 7.94 & 5.72 & 7.52 \\
TFC--TDF v3~\cite{tfctdfv3,melroformer} & 70.0M & 1 & 9.59 & 8.45 & 8.44 & 6.86 & 8.34 \\
SCNet~\cite{scnet} & 10.08M & 1 & 9.89 & 8.82 & 10.51 & 6.76 & 9.00 \\
SCNet-large~\cite{scnet} & 41.2M & 1 & 10.86 & \underline{9.49} & 10.98 & \underline{7.44} & 9.69 \\
\textbf{\smallmodel{}} (Ours) & 22.59M & 1 & \underline{11.72} & \textbf{9.59} & \underline{11.05} & \textbf{8.78} & \textbf{10.29} \\
\textbf{\basemodel{}} (Ours) & 97.23M & 1 & \textbf{12.10} & 8.81 & \textbf{11.08} & \textbf{8.78} & \underline{10.19} \\
\midrule
\end{tabular}
}
\parbox{\columnwidth}{\scriptsize
$N$ denotes the number of trained models used for multi-stem prediction.
Bold and underlined values denote the highest and second-highest reported
scores in each metric column within each model group, respectively;
ties receive the same emphasis.}
\end{table}

\subsection{Results}

We examine separation quality across single-stem and multi-stem
configurations, parameter requirements for four-stem prediction, and
vocals-only design ablations. Table~\ref{tab:baseline-comparison} presents
systems trained without extra data, grouped by output configuration.
Experiments with extra training data are presented separately in
Section~\ref{sec:extra-data}. For our evaluations, each stem score is
the median over the 50 MUSDB18-HQ test tracks, and the four-stem mean is
the arithmetic mean of these stem scores. The main proposed-model results
use the final EMA checkpoints and 2-s inference segments described in
Section~\ref{sec:experiments}; the ablation checkpoints are identified
separately below.

\subsection{Separation Performance}

Without extra training data, the single-stem version of \basemodel{}
achieves the highest reported four-stem mean among the complete results in
Table~\ref{tab:baseline-comparison}, at 10.63~dB. This is 0.61~dB above
BS-RoFormer, with higher vocals, drums, and other scores but a 1.38~dB
lower bass score. The single-stem version of \smallmodel{} reaches
9.82~dB. Increasing model size raises the mean by 0.81~dB and improves
vocals, bass, and other by 1.29, 1.02, and 1.11~dB, respectively, while
the smaller model scores 0.20~dB higher on drums. Thus, the higher mean
does not imply uniform gains across stems.

Joint prediction yields mean SDRs of 10.29~dB for the multi-stem version
of \smallmodel{} and 10.19~dB for the multi-stem version of \basemodel{}.
These are the two highest reported means in the multi-stem group.
The multi-stem version of \smallmodel{} exceeds
SCNet-large by 0.60~dB while using 22.59M rather than 41.2M parameters.
Relative to its single-stem version, it improves all four stem scores and
raises the mean by 0.47~dB. In contrast, the multi-stem version of
\basemodel{} scores 0.44~dB below its single-stem counterpart: drums
improve by 0.68~dB, but vocals, bass, and other decrease by 0.53, 1.39,
and 0.49~dB, respectively. Within the multi-stem configurations, the base
model scores higher on vocals and drums, ties on other, and scores lower
on bass than the small model. These results show that the observed
benefit of joint prediction depends on model size and target stem; test
scores alone do not establish overfitting as the cause.

Both single-stem and multi-stem configurations use one million optimizer
updates per model in the no-extra-data setting. Equal update counts do
not imply equal total training cost, since the single-stem configuration
requires four separately trained networks. The results establish the
reported performance with 2-s training crops and inference segments, but
do not isolate the effect of context length or show whether longer
segments would improve separation.

\subsection{Parameter Efficiency}

Parameter efficiency depends on whether size is measured per network or
for the complete four-stem system. In
Table~\ref{tab:baseline-comparison}, the single-stem version of
\smallmodel{} uses 22.52M parameters per network, or 90.08M across four
networks. It reaches 9.82~dB, compared with 10.02~dB for the 82.8M-per-network
BS-RoFormer configuration (331.2M in total). The multi-stem version of
\smallmodel{} requires only 22.59M parameters for all four stems and
reaches 10.29~dB. The corresponding base configurations require 388.68M
and 97.23M deployed parameters and reach 10.63 and 10.19~dB, respectively.
Thus, joint prediction reduces deployed parameter requirements for both
sizes, but does not consistently increase mean SDR.

\begin{figure}[t]
    \centering
    \includegraphics[width=\columnwidth]{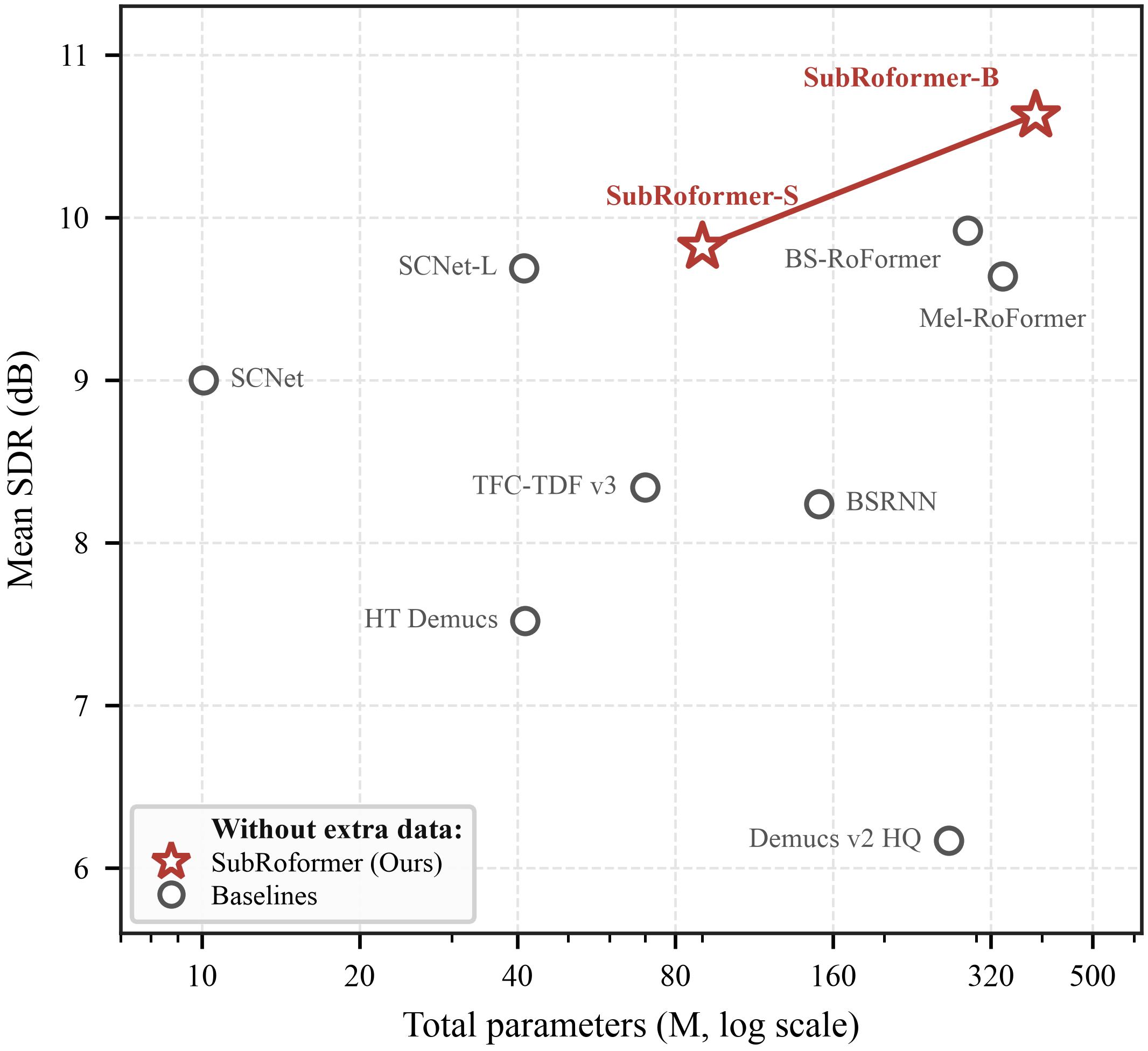}
    \caption{Mean test SDR (dB; higher is better) on MUSDB18-HQ versus
    total deployed parameters for selected systems trained without extra
    data. Single-stem systems, including both SubRoFormer sizes, use four
    networks; multi-stem baselines use one. The plotted BS-RoFormer and
    Mel-RoFormer points are six-layer configurations. Baseline results are
    drawn from~\cite{demucs,bsrnn,bsroformer,melroformer,htdemucs,tfctdfv3,scnet}.}
    \label{fig:parameter-efficiency}
\end{figure}

Figure~\ref{fig:parameter-efficiency} shows total deployed parameters for
selected no-extra-data systems, including the contextual six-layer
BS-RoFormer and Mel-RoFormer configurations rather than the configurations
listed in Table~\ref{tab:baseline-comparison}. In this plot, the
single-stem version of \smallmodel{} uses approximately 31\% and 27\% of
their respective total parameter counts, with a mean SDR 0.10~dB lower
than BS-RoFormer L6 and 0.18~dB higher than Mel-RoFormer L6. These
comparisons describe model-size trade-offs; they do not establish global
Pareto optimality or equivalent reductions in latency, memory use, or
training cost.

\begin{table}[t]
    \centering
    \caption{Parameter distribution of single-output variants with
    matched separators.}
    \label{tab:training-vram-parameters}
    \footnotesize
    \setlength{\tabcolsep}{3pt}
    \resizebox{\columnwidth}{!}{%
    \begin{tabular}{@{}lrrr@{}}
        \toprule
        & Total (M) & Representation + mask (M)
        & Separator (M) \\
        \midrule
        BS-RoFormer L6 & 72.19
        & 50.87 (70.47\%) & 21.32 (29.53\%) \\
        \midrule
        Mel-RoFormer L6 & 84.19
        & 62.87 (74.68\%) & 21.32 (25.32\%) \\
        \midrule
        SubRoFormer L4+2 & \textbf{24.08}
        & \textbf{2.76 (11.45\%)} & 21.32 (88.55\%) \\
        \bottomrule
    \end{tabular}%
    }
    \par\smallskip
    \begin{minipage}{\columnwidth}
        \footnotesize
        Parameter counts are in millions, with group shares in parentheses.
        All variants use six identical width-384 BS-RoFormer blocks and
        one terminal normalization. SubRoFormer arranges the blocks as
        one pre block, four central blocks, and one post block.
    \end{minipage}
\end{table}

Table~\ref{tab:training-vram-parameters} compares how single-output
variants allocate parameters between representation and mask prediction
and the separator. To hold separator capacity fixed, all three variants
use six identical width-384 BS-RoFormer
axial blocks with 12 attention heads, head dimension 32, feed-forward
ratio four, and one terminal RMSNorm. SubRoFormer arranges these blocks
as $1+4+1$ with temporal patch size four; its pre/post blocks are replaced
by full-width BS-RoFormer blocks. The Mel-RoFormer variant replaces its
native separator with the same BS-style stack and terminal normalization.
BS-RoFormer and Mel-RoFormer~\cite{bsroformer,melroformer} retain their
constructor-default band splits and mask heads from the ZFTurbo
implementation~\cite{msst}. These variants are used only for the
parameter-distribution comparison and are distinct from the models used
for the separation-quality results.

Parameter counts use unique registered model parameters. Representation
includes input/output projections, temporal down/up layers, skip fusion,
and mask prediction; separator includes all six axial blocks and the
terminal normalization. Each separator has exactly 21,321,776 registered
parameters, including 32 frozen rotary parameters. SubRoFormer allocates
2.76M parameters (11.45\%) to representation and mask prediction,
compared with 50.87M (70.47\%) for BS-RoFormer and 62.87M (74.68\%) for
Mel-RoFormer. Its fixed band-splitting filters have no learned parameters.
The separator accounts for 88.55\% of SubRoFormer's 24.08M parameters,
versus 29.53\% of BS-RoFormer's 72.19M and 25.32\% of Mel-RoFormer's
84.19M. With separator counts held equal, the total-parameter differences
come from the representation and mask-prediction modules.

\subsection{Vocals Ablations}

Table~\ref{tab:vocals-ablations} examines frontend, temporal-resolution,
and fusion choices using the single-stem version of \smallmodel{}.
The reference and ablation scores use 400k-update EMA checkpoints and
are separate from the one-million-update main results. The 10.88-dB
reference comes from a separate preliminary run, and each configuration
has one seed; differences are therefore descriptive rather than
statistically established effects. The reproduced six-layer BS-RoFormer
and Mel-RoFormer scores use one-million-update checkpoints and a
different inference segment length, so they serve only as contextual
references.

For frontend representation (A1), replacing local subband analysis with
global-STFT band-split or mel band-split features yields 9.43 and
8.88~dB, respectively, below the 10.88-dB reference. These replacements
use linear projection into the backbone rather than reproducing the
original models' complete frontends. For subband decomposition (A2),
the 64-subband configuration scores 9.57~dB, 1.31~dB below the
128-subband reference, while matching its 100-frame/s local-feature and
25-frame/s backbone grids. This comparison holds the temporal rates
fixed, but does not isolate subband count from the accompanying changes
in subband bandwidth and sampling rate.

Reducing the temporal patch factor from four to one or two (A3) yields
11.02 and 10.94~dB, respectively, improvements of 0.14 and 0.06~dB.
These smaller factors retain more temporal positions in the backbone.
The scores suggest a quality--sequence-length trade-off, but no runtime
measurements or repeated-seed estimates are provided to quantify its
computational cost or robustness.

For skip fusion (A4), the GEGLU reference scores above no skip fusion
(10.45~dB), additive fusion (10.77~dB), multiplicative fusion
(10.56~dB), and linear concatenation (10.57~dB). For local feature
analysis (A5), CNN analysis and direct subband samples reach 10.66 and
10.43~dB, respectively, below local STFT analysis. The
GELU~\cite{gelu} feed-forward
variant (A6) reaches 10.55~dB, 0.33~dB below the reference. This last
comparison concerns the feed-forward activation, not the GEGLU
skip-fusion operator.

Overall, local subband STFT analysis has the highest score among the
tested frontend variants, while smaller temporal patch factors yield
the highest vocals scores in this ablation set. These single-seed,
vocals-only observations support further investigation of the design
choices, but do not establish their causal contributions to the
four-stem results.

\begin{table}[t]
\caption{Vocals-only ablation results on MUSDB18-HQ.}
\label{tab:vocals-ablations}
\centering
\normalsize
\setlength{\tabcolsep}{4pt}
\setlength{\aboverulesep}{0.25ex}
\setlength{\belowrulesep}{0.25ex}
\renewcommand{\arraystretch}{1.0}
\begin{tabular*}{\columnwidth}{@{}p{0.60\columnwidth}@{\extracolsep{\fill}}rr@{}}
\toprule
Variant & SDR $\uparrow$ & $\Delta$SDR $\uparrow$ \\
\midrule
\smallmodel{} (400k updates) & 10.88 & $0.00$ \\
\midrule
BS-RoFormer L6$^\ast$~\cite{bsroformer} & 10.18 & -- \\
Mel-RoFormer L6$^\ast$~\cite{melroformer} & 10.08 & -- \\
\midrule
\multicolumn{3}{@{}l}{\textit{A1: Front-end representation}} \\
\quad Global-STFT band-split & 9.43 & $-1.45$ \\
\quad Mel band-split & 8.88 & $-2.00$ \\
\midrule
\multicolumn{3}{@{}l}{\textit{A2: Subband decomposition}} \\
\quad 64 subbands & 9.57 & $-1.31$ \\
\midrule
\multicolumn{3}{@{}l}{\textit{A3: Temporal patching}} \\
\quad Temporal patch factor 1 & \textbf{11.02} & \textbf{$+0.14$} \\
\quad Temporal patch factor 2 & \underline{10.94} & $+0.06$ \\
\midrule
\multicolumn{3}{@{}l}{\textit{A4: Skip-fusion operator}} \\
\quad No skip fusion & 10.45 & $-0.43$ \\
\quad Additive skip fusion & 10.77 & $-0.11$ \\
\quad Multiplicative skip fusion & 10.56 & $-0.32$ \\
\quad Linear concatenation fusion & 10.57 & $-0.31$ \\
\midrule
\multicolumn{3}{@{}l}{\textit{A5: Local feature analysis}} \\
\quad CNN analysis & 10.66 & $-0.22$ \\
\quad Direct subband samples & 10.43 & $-0.45$ \\
\midrule
\multicolumn{3}{@{}l}{\textit{A6: Feed-forward activation}} \\
\quad GELU feed-forward network & 10.55 & $-0.33$ \\
\bottomrule
\end{tabular*}
\end{table}

\subsection{Extra Data}
\label{sec:extra-data}

To examine performance with additional training data, we supplement the 100
MUSDB18-HQ training songs with an internal dataset of 500 additional songs.
The extra-data four-stem \basemodel\ in
Table~\ref{tab:baseline-comparison-extra} is trained for 1.4 million optimizer
updates on eight NVIDIA GeForce RTX 4090 GPUs with a per-GPU batch size of two,
giving a global batch size of 16.

\begin{table}[t]
\caption{MUSDB18-HQ metrics for systems trained with extra data and a
contextual system with unknown training data.
Metrics for research baselines are taken from their original papers.}
\label{tab:baseline-comparison-extra}
\centering
\scriptsize
\setlength{\tabcolsep}{1.5pt}
\renewcommand{\arraystretch}{1.12}
\resizebox{\columnwidth}{!}{%
\begin{tabular}{@{}lcccccccc@{}}
\toprule
Model & Params. & $N$ & Extra & Voc. $\uparrow$ & Bass $\uparrow$ & Dr. $\uparrow$ &
Oth. $\uparrow$ & Avg. $\uparrow$ \\
\midrule
\multicolumn{9}{c}{\textit{Single-stem models}} \\
\midrule
HT Demucs (FT)~\cite{htdemucs} & 41.4M & 4 & 800 & 9.20 & \underline{10.39} & 10.08 & 6.32 & \underline{9.00} \\
BSRNN~\cite{bsrnn} & 37.6M & 4 & 1750 & \underline{10.47} & 8.16 & \underline{10.15} & \underline{7.08} & 8.97 \\
BS-RoFormer L12~\cite{bsroformer} & 93.4M & 3 & 450 & \textbf{12.72} & \textbf{13.32} & \textbf{12.91} & \textbf{9.01} & \textbf{11.99} \\
\midrule
\multicolumn{9}{c}{\textit{Multi-stem models}} \\
\midrule
Hybrid Demucs~\cite{hybriddemucs,htdemucs} & 83.64M & 1 & 800 & 8.75 & 9.13 & 9.31 & 6.18 & 8.34 \\
SCNet~\cite{scnet} & 10.08M & 1 & 235 & 10.17 & 9.21 & 10.78 & 6.84 & 9.25 \\
SCNet-large~\cite{scnet} & 41.2M & 1 & 235 & \underline{11.10} & \underline{9.86} & \underline{11.23} & \underline{7.51} & \underline{9.92} \\
\textbf{\basemodel{}} (Ours) & 97.23M & 1 & 500 & \textbf{12.69} & \textbf{11.50} & \textbf{12.23} & \textbf{9.94} & \textbf{11.59} \\
\midrule
\multicolumn{9}{c}{\textit{Commercial systems}} \\
\midrule
ACE Studio~\cite{acestudio} & -- & -- & -- & 11.67 & 11.83 & 10.93 & 9.26 & 10.92 \\
\midrule
\end{tabular}
}
\parbox{\columnwidth}{\scriptsize
$N$ denotes the number of trained models used for multi-stem prediction.
Bold and underlined values denote the highest and second-highest reported
scores in each metric column within each model group, respectively;
ties receive the same emphasis.
BS-RoFormer L12 uses three 93.4M target-specific networks to cover the four stems.}
\end{table}

With 500 extra training songs, the multi-stem version of \basemodel{}
reaches 11.59~dB in Table~\ref{tab:baseline-comparison-extra}, with
12.69, 11.50, 12.23, and 9.94~dB for vocals, bass, drums, and other,
respectively. Its mean is 1.40~dB above the no-extra-data multi-stem
version, with higher scores on every stem. However, the extra-data run
also uses a longer training schedule and a different global batch size;
the gain cannot be attributed to additional data alone.

BS-RoFormer L12 retains the higher extra-data mean of 11.99~dB, compared
with 11.59~dB for the multi-stem version of \basemodel{}. The latter
scores 0.93~dB higher on other but lower on vocals, bass, and drums.
BS-RoFormer L12 uses three 93.4M target-specific networks, totaling
280.2M parameters, and obtains other as the mixture residual. The
multi-stem version of \basemodel{} instead predicts all four stems with
one 97.23M network, approximately 35\% of that deployed parameter count.
The training data (500 versus 450 extra songs) and checkpoint generations
are not matched, so this is a descriptive comparison of separation
quality and deployed model size, not a controlled architectural study.

\subsubsection{Commercial-System Comparison}

Table~\ref{tab:baseline-comparison-extra} includes ACE Studio's stem
splitter~\cite{acestudio} as a contextual reference. The supplied
four-stem exports were obtained with the default splitter settings.
Their reported mean SDR is 10.92~dB in the local evaluation over all
50 MUSDB18-HQ test songs. The no-extra-data multi-stem versions of \smallmodel{} and
\basemodel{} score 10.29 and 10.19~dB, respectively, below this reference.
With extra data, the multi-stem version of \basemodel{} reaches
11.59~dB, a 0.67~dB higher mean. Its vocals, drums, and other scores are
higher by 1.02, 1.30, and 0.68~dB, respectively, whereas ACE Studio
scores 0.33~dB higher on bass.

This comparison concerns the evaluated outputs rather than matched
architectures or training conditions. ACE Studio's scores use lossy
Opus exports, and its model/version, training data, input-mixture
provenance, and parameter count are unknown. Although aggregation follows
our local evaluation convention, inference and audio-format protocols
are not fully matched. The result therefore contextualizes separation
quality for one commercial system; it does not establish general
superiority over commercial systems or greater parameter efficiency.

\section{Conclusion}
We presented \model, a target-specific MSS architecture that combines an
overlapping waveform subband bank, local complex time--frequency analysis, and
a multiscale axial RoFormer.  The design predicts complex masks at 100~Hz while
placing its widest sequence model at 25~Hz.  The 97.17M configuration reaches
a preliminary 10.63-dB four-stem average on the supplied MUSDB18-HQ record,
0.61~dB above the strongest complete author-reported public-data BS-RoFormer
result and 0.99~dB above the reported public-data Mel-RoFormer result.  The
comparison is favorable across all four stems against Mel-RoFormer, while
BS-RoFormer retains a 1.38-dB advantage on bass.  In a separate vocals-only
study, two global-STFT replacements score below the 10.88-dB reference, while
temporal patch factors one and two obtain the two highest ablation scores of
11.02 and 10.94~dB.  These single-seed observations are directional rather
than causal.  Versioned evaluation artifacts, repeated seeds, efficiency
measurements, and parameter-matched all-stem ablations are required to convert
this promising draft result into a robust journal claim.

\bibliographystyle{IEEEtran}
\bibliography{ref}

@book{oppenheimdsp,
  author       = {Alan V. Oppenheim and Ronald W. Schafer},
  title        = {Discrete-Time Signal Processing},
  edition      = {3},
  publisher    = {Pearson Prentice Hall},
  year         = {2010}
}

@article{melscale,
  author       = {S. S. Stevens and J. Volkmann and E. B. Newman},
  title        = {A Scale for the Measurement of the Psychological
                  Magnitude Pitch},
  journal      = {The Journal of the Acoustical Society of America},
  volume       = {8},
  number       = {3},
  pages        = {185--190},
  year         = {1937},
  doi          = {10.1121/1.1915893}
}

@article{erbscale,
  author       = {Brian R. Glasberg and Brian C. J. Moore},
  title        = {Derivation of Auditory Filter Shapes from Notched-Noise
                  Data},
  journal      = {Hearing Research},
  volume       = {47},
  number       = {1--2},
  pages        = {103--138},
  year         = {1990},
  doi          = {10.1016/0378-5955(90)90170-T}
}

@inproceedings{mdct,
  author       = {J. P. Princen and A. W. Johnson and A. B. Bradley},
  title        = {Subband/Transform Coding Using Filter Bank Designs Based
                  on Time Domain Aliasing Cancellation},
  booktitle    = {IEEE International Conference on Acoustics, Speech, and
                  Signal Processing},
  volume       = {12},
  pages        = {2161--2164},
  publisher    = {IEEE},
  year         = {1987},
  doi          = {10.1109/ICASSP.1987.1169405}
}

@article{waveletpackets,
  author       = {Ronald R. Coifman and Mladen Victor Wickerhauser},
  title        = {Entropy-Based Algorithms for Best Basis Selection},
  journal      = {IEEE Transactions on Information Theory},
  volume       = {38},
  number       = {2},
  pages        = {713--718},
  year         = {1992},
  doi          = {10.1109/18.119732}
}

@inproceedings{sgdr,
  author       = {Ilya Loshchilov and Frank Hutter},
  title        = {{SGDR}: Stochastic Gradient Descent with Warm Restarts},
  booktitle    = {International Conference on Learning Representations},
  year         = {2017},
  url          = {https://arxiv.org/abs/1608.03983}
}

@article{gelu,
  author       = {Dan Hendrycks and Kevin Gimpel},
  title        = {Gaussian Error Linear Units ({GELUs})},
  journal      = {arXiv preprint arXiv:1606.08415},
  year         = {2016},
  doi          = {10.48550/arXiv.1606.08415}
}

@misc{acestudio,
  author       = {{ACE Studio}},
  title        = {{AI} Stem Splitter},
  url          = {https://acestudio.ai/stem-splitter/},
  note         = {Accessed: Sep. 8, 2026}
}

@misc{musdb,
  author       = {Zafar Rafii and Antoine Liutkus and
                  Fabian-Robert St{\"o}ter and Stylianos Ioannis Mimilakis and
                  Rachel Bittner},
  title        = {The {MUSDB18} Corpus for Music Separation},
  month        = dec,
  year         = {2017},
  doi          = {10.5281/zenodo.1117372},
  url          = {https://doi.org/10.5281/zenodo.1117372}
}

@misc{musdbhq,
  author       = {Zafar Rafii and Antoine Liutkus and
                  Fabian-Robert St{\"o}ter and Stylianos Ioannis Mimilakis and
                  Rachel Bittner},
  title        = {{MUSDB18-HQ}: An Uncompressed Version of {MUSDB18}},
  month        = aug,
  year         = {2019},
  doi          = {10.5281/zenodo.3338373},
  url          = {https://doi.org/10.5281/zenodo.3338373}
}

@misc{msst,
  author       = {{ZFTurbo}},
  title        = {Music Source Separation Training},
  howpublished = {GitHub repository},
  url          = {https://github.com/ZFTurbo/Music-Source-Separation-Training},
  note         = {Accessed: Aug. 19, 2026}
}

@inproceedings{sisec2018,
  author       = {Fabian-Robert St{\"o}ter and Antoine Liutkus and Nobutaka Ito},
  title        = {The 2018 Signal Separation Evaluation Campaign},
  booktitle    = {Latent Variable Analysis and Signal Separation},
  series       = {Lecture Notes in Computer Science},
  volume       = {10891},
  pages        = {293--305},
  publisher    = {Springer},
  year         = {2018},
  doi          = {10.1007/978-3-319-93764-9_28}
}

@book{bregman1990,
  author       = {Albert S. Bregman},
  title        = {Auditory Scene Analysis: The Perceptual Organization of
                  Sound},
  publisher    = {MIT Press},
  address      = {Cambridge, MA, USA},
  year         = {1990},
  doi          = {10.7551/mitpress/1486.001.0001}
}

@article{bell1995,
  author       = {Anthony J. Bell and Terrence J. Sejnowski},
  title        = {An Information-Maximization Approach to Blind Separation
                  and Blind Deconvolution},
  journal      = {Neural Computation},
  volume       = {7},
  number       = {6},
  pages        = {1129--1159},
  year         = {1995},
  doi          = {10.1162/neco.1995.7.6.1129}
}

@article{smaragdis1998,
  author       = {Paris Smaragdis},
  title        = {Blind Separation of Convolved Mixtures in the Frequency
                  Domain},
  journal      = {Neurocomputing},
  volume       = {22},
  number       = {1--3},
  pages        = {21--34},
  year         = {1998},
  doi          = {10.1016/S0925-2312(98)00047-2}
}

@inproceedings{virtanenklapuri2002,
  author       = {Tuomas Virtanen and Anssi Klapuri},
  title        = {Separation of Harmonic Sounds Using Linear Models for the
                  Overtone Series},
  booktitle    = {IEEE International Conference on Acoustics, Speech and
                  Signal Processing},
  volume       = {2},
  pages        = {II-1757--II-1760},
  year         = {2002},
  doi          = {10.1109/ICASSP.2002.5744962}
}

@article{virtanennmf,
  author       = {Tuomas Virtanen},
  title        = {Monaural Sound Source Separation by Nonnegative Matrix
                  Factorization With Temporal Continuity and Sparseness
                  Criteria},
  journal      = {IEEE Transactions on Audio, Speech, and Language Processing},
  volume       = {15},
  number       = {3},
  pages        = {1066--1074},
  year         = {2007},
  doi          = {10.1109/TASL.2006.885253}
}

@article{huangmonaural,
  author       = {Po-Sen Huang and Minje Kim and Mark Hasegawa-Johnson and
                  Paris Smaragdis},
  title        = {Joint Optimization of Masks and Deep Recurrent Neural
                  Networks for Monaural Source Separation},
  journal      = {IEEE/ACM Transactions on Audio, Speech, and Language
                  Processing},
  volume       = {23},
  number       = {12},
  pages        = {2136--2147},
  year         = {2015},
  doi          = {10.1109/TASLP.2015.2468583}
}

@article{rafiioverview,
  author       = {Zafar Rafii and Antoine Liutkus and Fabian-Robert St{\"o}ter
                  and Stylianos Ioannis Mimilakis and Derry FitzGerald and
                  Bryan Pardo},
  title        = {An Overview of Lead and Accompaniment Separation in Music},
  journal      = {IEEE/ACM Transactions on Audio, Speech, and Language
                  Processing},
  volume       = {26},
  number       = {8},
  pages        = {1307--1335},
  year         = {2018},
  doi          = {10.1109/TASLP.2018.2825440}
}

@article{cano2019mss,
  author       = {Estefan{\'i}a Cano and Derry FitzGerald and Antoine Liutkus
                  and Mark D. Plumbley and Fabian-Robert St{\"o}ter},
  title        = {Musical Source Separation: An Introduction},
  journal      = {IEEE Signal Processing Magazine},
  volume       = {36},
  number       = {1},
  pages        = {31--40},
  year         = {2019},
  doi          = {10.1109/MSP.2018.2874719}
}

@inproceedings{waveunet,
  author       = {Daniel Stoller and Sebastian Ewert and Simon Dixon},
  title        = {{Wave-U-Net}: A Multi-Scale Neural Network for End-to-End
                  Audio Source Separation},
  booktitle    = {Proceedings of the 19th International Society for Music
                  Information Retrieval Conference},
  pages        = {334--340},
  year         = {2018},
  url          = {https://archives.ismir.net/ismir2018/paper/000205.pdf}
}

@article{demucs,
  author       = {Alexandre D{\'e}fossez and Nicolas Usunier and
                  L{\'e}on Bottou and Francis Bach},
  title        = {Music Source Separation in the Waveform Domain},
  journal      = {arXiv preprint arXiv:1911.13254},
  year         = {2019},
  doi          = {10.48550/arXiv.1911.13254}
}

@inproceedings{hybriddemucs,
  author       = {Alexandre D{\'e}fossez},
  title        = {Hybrid Spectrogram and Waveform Source Separation},
  booktitle    = {Proceedings of the ISMIR 2021 Workshop on Music Source
                  Separation},
  pages        = {1--11},
  year         = {2021},
  url          = {https://mdx-workshop.github.io/proceedings/defossez.pdf}
}

@inproceedings{htdemucs,
  author       = {Simon Rouard and Francisco Massa and
                  Alexandre D{\'e}fossez},
  title        = {Hybrid Transformers for Music Source Separation},
  booktitle    = {IEEE International Conference on Acoustics, Speech and
                  Signal Processing},
  pages        = {1--5},
  year         = {2023},
  doi          = {10.1109/ICASSP49357.2023.10096956}
}

@article{openunmix,
  author       = {Fabian-Robert St{\"o}ter and Stefan Uhlich and
                  Antoine Liutkus and Yuki Mitsufuji},
  title        = {{Open-Unmix}---A Reference Implementation for Music Source
                  Separation},
  journal      = {Journal of Open Source Software},
  volume       = {4},
  number       = {41},
  pages        = {1667},
  year         = {2019},
  doi          = {10.21105/joss.01667}
}

@article{bsrnn,
  author       = {Yi Luo and Jianwei Yu},
  title        = {Music Source Separation With Band-Split {RNN}},
  journal      = {IEEE/ACM Transactions on Audio, Speech, and Language
                  Processing},
  volume       = {31},
  pages        = {1893--1901},
  year         = {2023},
  doi          = {10.1109/TASLP.2023.3271145}
}

@inproceedings{bsroformer,
  author       = {Wei-Tsung Lu and Ju-Chiang Wang and Qiuqiang Kong and
                  Yun-Ning Hung},
  title        = {Music Source Separation With Band-Split {RoPE} Transformer},
  booktitle    = {IEEE International Conference on Acoustics, Speech and
                  Signal Processing},
  pages        = {481--485},
  publisher    = {IEEE},
  year         = {2024},
  doi          = {10.1109/ICASSP48485.2024.10446843}
}

@article{melroformer,
  author       = {Ju-Chiang Wang and Wei-Tsung Lu and Minz Won},
  title        = {Mel-Band {RoFormer} for Music Source Separation},
  journal      = {arXiv preprint arXiv:2310.01809},
  year         = {2023},
  doi          = {10.48550/arXiv.2310.01809}
}

@inproceedings{scnet,
  author       = {Weinan Tong and Jiaxu Zhu and Jun Chen and Shiyin Kang and
                  Tao Jiang and Yang Li and Zhiyong Wu and Helen Meng},
  title        = {{SCNet}: Sparse Compression Network for Music Source
                  Separation},
  booktitle    = {IEEE International Conference on Acoustics, Speech and
                  Signal Processing},
  pages        = {1276--1280},
  publisher    = {IEEE},
  year         = {2024},
  doi          = {10.1109/ICASSP48485.2024.10446651}
}

@inproceedings{dttnet,
  author       = {Junyu Chen and Susmitha Vekkot and Pancham Shukla},
  title        = {Music Source Separation Based on a Lightweight Deep
                  Learning Framework ({DTTNet}: Dual-Path {TFC-TDF UNet})},
  booktitle    = {IEEE International Conference on Acoustics, Speech and
                  Signal Processing},
  pages        = {656--660},
  publisher    = {IEEE},
  year         = {2024},
  doi          = {10.1109/ICASSP48485.2024.10448020},
  url          = {https://arxiv.org/abs/2309.08684}
}

@article{tfctdfv3,
  author       = {Minseok Kim and Jun Hyung Lee and Soonyoung Jung},
  title        = {Sound Demixing Challenge 2023 Music Demixing Track
                  Technical Report: {TFC-TDF-UNet v3}},
  journal      = {arXiv preprint arXiv:2306.09382},
  year         = {2023},
  doi          = {10.48550/arXiv.2306.09382}
}

@article{roformer,
  author       = {Jianlin Su and Murtadha Ahmed and Yu Lu and
                  Shengfeng Pan and Wen Bo and Yunfeng Liu},
  title        = {{RoFormer}: Enhanced Transformer with Rotary Position
                  Embedding},
  journal      = {Neurocomputing},
  volume       = {568},
  pages        = {127063},
  year         = {2024},
  doi          = {10.1016/j.neucom.2023.127063}
}

@inproceedings{transformer,
  author       = {Ashish Vaswani and Noam Shazeer and Niki Parmar and
                  Jakob Uszkoreit and Llion Jones and Aidan N. Gomez and
                  {\L}ukasz Kaiser and Illia Polosukhin},
  title        = {Attention Is All You Need},
  booktitle    = {Advances in Neural Information Processing Systems},
  volume       = {30},
  pages        = {5998--6008},
  year         = {2017}
}

@inproceedings{rmsnorm,
  author       = {Biao Zhang and Rico Sennrich},
  title        = {Root Mean Square Layer Normalization},
  booktitle    = {Advances in Neural Information Processing Systems},
  volume       = {32},
  pages        = {12360--12371},
  year         = {2019}
}

@article{swiglu,
  author       = {Noam Shazeer},
  title        = {{GLU} Variants Improve Transformer},
  journal      = {arXiv preprint arXiv:2002.05202},
  year         = {2020},
  doi          = {10.48550/arXiv.2002.05202}
}

@inproceedings{convnextv2,
  author       = {Sanghyun Woo and Shoubhik Debnath and Ronghang Hu and
                  Xinlei Chen and Zhuang Liu and In So Kweon and Saining Xie},
  title        = {{ConvNeXt V2}: Co-Designing and Scaling {ConvNets} with
                  Masked Autoencoders},
  booktitle    = {IEEE/CVF Conference on Computer Vision and Pattern
                  Recognition},
  pages        = {16133--16142},
  publisher    = {IEEE},
  year         = {2023},
  doi          = {10.1109/CVPR52729.2023.01548}
}

@inproceedings{parallelwavegan,
  author       = {Ryuichi Yamamoto and Eunwoo Song and Jae-Min Kim},
  title        = {Parallel {WaveGAN}: A Fast Waveform Generation Model Based
                  on Generative Adversarial Networks with Multi-Resolution
                  Spectrogram},
  booktitle    = {IEEE International Conference on Acoustics, Speech and
                  Signal Processing},
  pages        = {6199--6203},
  publisher    = {IEEE},
  year         = {2020},
  doi          = {10.1109/ICASSP40776.2020.9053795}
}

@article{bsseval,
  author       = {Emmanuel Vincent and R{\'e}mi Gribonval and
                  C{\'e}dric F{\'e}votte},
  title        = {Performance Measurement in Blind Audio Source Separation},
  journal      = {IEEE Transactions on Audio, Speech, and Language Processing},
  volume       = {14},
  number       = {4},
  pages        = {1462--1469},
  year         = {2006},
  doi          = {10.1109/TSA.2005.858005}
}

@inproceedings{cws,
  author       = {Haohe Liu and Lei Xie and Jian Wu and Geng Yang},
  title        = {Channel-wise Subband Input for Better Voice and
                  Accompaniment Separation on High Resolution Music},
  booktitle    = {Proceedings of Interspeech},
  pages        = {1241--1245},
  year         = {2020},
  doi          = {10.21437/Interspeech.2020-2555}
}

@inproceedings{cwspresunet,
  author       = {Haohe Liu and Qiuqiang Kong and Jiafeng Liu},
  title        = {{CWS-PResUNet}: Music Source Separation with Channel-wise
                  Subband Phase-aware {ResUNet}},
  booktitle    = {Proceedings of the ISMIR 2021 Workshop on Music Source
                  Separation},
  pages        = {1--5},
  year         = {2021},
  url          = {https://mdx-workshop.github.io/proceedings/liu.pdf}
}

@inproceedings{xumxslicq,
  author       = {Sevag Hanssian},
  title        = {Music Demixing with the {sliCQ} Transform},
  booktitle    = {Proceedings of the ISMIR 2021 Workshop on Music Source
                  Separation},
  pages        = {1--3},
  year         = {2021},
  url          = {https://mdx-workshop.github.io/proceedings/hanssian.pdf}
}

@inproceedings{xumxslicqv2,
  author       = {Sevag Hanssian},
  title        = {Better Music Demixing with the {sliCQ} Transform},
  booktitle    = {Proceedings of the Cadenza Workshop},
  year         = {2023},
  url          = {https://cadenzachallenge.org/cadenza2023-workshop/papers/cad1_23_hanssian_paper.pdf}
}

@article{multiratetutorial,
  author       = {P. P. Vaidyanathan},
  title        = {Multirate Digital Filters, Filter Banks, Polyphase Networks,
                  and Applications: A Tutorial},
  journal      = {Proceedings of the IEEE},
  volume       = {78},
  number       = {1},
  pages        = {56--93},
  year         = {1990},
  doi          = {10.1109/5.52200}
}

@article{stftanalysis,
  author       = {Jont B. Allen and Lawrence R. Rabiner},
  title        = {A Unified Approach to Short-Time Fourier Analysis and
                  Synthesis},
  journal      = {Proceedings of the IEEE},
  volume       = {65},
  number       = {11},
  pages        = {1558--1564},
  year         = {1977},
  doi          = {10.1109/PROC.1977.10770}
}

@article{browncqt,
  author       = {Judith C. Brown},
  title        = {Calculation of a Constant {Q} Spectral Transform},
  journal      = {The Journal of the Acoustical Society of America},
  volume       = {89},
  number       = {1},
  pages        = {425--434},
  year         = {1991},
  doi          = {10.1121/1.400476},
  url          = {https://www.ee.columbia.edu/~dpwe/papers/Brown91-cqt.pdf}
}

@article{invertiblecqt,
  author       = {Nicki Holighaus and Monika D{\"o}rfler and Gino Angelo
                  Velasco and Thomas Grill},
  title        = {A Framework for Invertible, Real-Time Constant-{Q}
                  Transforms},
  journal      = {IEEE Transactions on Audio, Speech, and Language Processing},
  volume       = {21},
  number       = {4},
  pages        = {775--785},
  year         = {2013},
  doi          = {10.1109/TASL.2012.2234114}
}

@article{nsgabor,
  author       = {Peter Balazs and Monika D{\"o}rfler and Florent Jaillet and
                  Nicki Holighaus and Gino Angelo Velasco},
  title        = {Theory, Implementation and Applications of Nonstationary
                  {Gabor} Frames},
  journal      = {Journal of Computational and Applied Mathematics},
  volume       = {236},
  number       = {6},
  pages        = {1481--1496},
  year         = {2011},
  doi          = {10.1016/j.cam.2011.09.011}
}

@inproceedings{singingunet,
  author       = {Andreas Jansson and Eric Humphrey and Nicola Montecchio and
                  Rachel Bittner and Aparna Kumar and Tillman Weyde},
  title        = {Singing Voice Separation with Deep {U-Net} Convolutional
                  Networks},
  booktitle    = {Proceedings of the 18th International Society for Music
                  Information Retrieval Conference},
  pages        = {745--751},
  year         = {2017},
  url          = {https://archives.ismir.net/ismir2017/paper/000171.pdf}
}

@article{spleeter,
  author       = {Romain Hennequin and Anis Khlif and Felix Voituret and
                  Manuel Moussallam},
  title        = {Spleeter: A Fast and Efficient Music Source Separation Tool
                  with Pre-trained Models},
  journal      = {Journal of Open Source Software},
  volume       = {5},
  number       = {50},
  pages        = {2154},
  year         = {2020},
  doi          = {10.21105/joss.02154}
}

@inproceedings{mmdenselstm,
  author       = {Naoya Takahashi and Nabarun Goswami and Yuki Mitsufuji},
  title        = {{MMDenseLSTM}: An Efficient Combination of Convolutional and
                  Recurrent Neural Networks for Audio Source Separation},
  booktitle    = {16th International Workshop on Acoustic Signal Enhancement},
  pages        = {106--110},
  publisher    = {IEEE},
  year         = {2018},
  doi          = {10.1109/IWAENC.2018.8521383},
  url          = {https://arxiv.org/abs/1805.02410}
}

@article{d3net,
  author       = {Naoya Takahashi and Yuki Mitsufuji},
  title        = {{D3Net}: Densely Connected Multidilated {DenseNet} for Music
                  Source Separation},
  journal      = {arXiv preprint arXiv:2010.01733},
  year         = {2020},
  doi          = {10.48550/arXiv.2010.01733}
}

@inproceedings{xumx,
  author       = {Ryosuke Sawata and Stefan Uhlich and Shusuke Takahashi and
                  Yuki Mitsufuji},
  title        = {All for One and One for All: Improving Music Separation by
                  Bridging Networks},
  booktitle    = {IEEE International Conference on Acoustics, Speech and
                  Signal Processing},
  pages        = {51--55},
  publisher    = {IEEE},
  year         = {2021},
  doi          = {10.1109/ICASSP39728.2021.9414044},
  url          = {https://arxiv.org/abs/2010.04228}
}

@inproceedings{kuielabmdx,
  author       = {Minseok Kim and Woosung Choi and Jaehwa Chung and Daewon Lee
                  and Soonyoung Jung},
  title        = {{KUIELab-MDX-Net}: A Two-Stream Neural Network for Music
                  Demixing},
  booktitle    = {Proceedings of the ISMIR 2021 Workshop on Music Source
                  Separation},
  year         = {2021},
  url          = {https://arxiv.org/abs/2111.12203}
}

@inproceedings{phaseresunet,
  author       = {Qiuqiang Kong and Yin Cao and Haohe Liu and Keunwoo Choi and
                  Yuxuan Wang},
  title        = {Decoupling Magnitude and Phase Estimation with Deep
                  {ResUNet} for Music Source Separation},
  booktitle    = {Proceedings of the 22nd International Society for Music
                  Information Retrieval Conference},
  pages        = {342--349},
  year         = {2021},
  url          = {https://arxiv.org/abs/2109.05418}
}

@inproceedings{lasaft,
  author       = {Woosung Choi and Minseok Kim and Jaehwa Chung and Soonyoung
                  Jung},
  title        = {{LaSAFT}: Latent Source Attentive Frequency Transformation
                  for Conditioned Source Separation},
  booktitle    = {IEEE International Conference on Acoustics, Speech and
                  Signal Processing},
  pages        = {171--175},
  year         = {2021},
  doi          = {10.1109/ICASSP39728.2021.9413896},
  url          = {https://arxiv.org/abs/2010.11631}
}

@article{convtasnet,
  author       = {Yi Luo and Nima Mesgarani},
  title        = {{Conv-TasNet}: Surpassing Ideal Time--Frequency Magnitude
                  Masking for Speech Separation},
  journal      = {IEEE/ACM Transactions on Audio, Speech, and Language
                  Processing},
  volume       = {27},
  number       = {8},
  pages        = {1256--1266},
  year         = {2019},
  doi          = {10.1109/TASLP.2019.2915167},
  url          = {https://arxiv.org/abs/1809.07454}
}

@inproceedings{dprnn,
  author       = {Yi Luo and Zhuo Chen and Takuya Yoshioka},
  title        = {Dual-Path {RNN}: Efficient Long Sequence Modeling for
                  Time-Domain Single-Channel Speech Separation},
  booktitle    = {IEEE International Conference on Acoustics, Speech and
                  Signal Processing},
  pages        = {46--50},
  year         = {2020},
  doi          = {10.1109/ICASSP40776.2020.9054266},
  url          = {https://arxiv.org/abs/1910.06379}
}

@article{complexmask,
  author       = {Donald S. Williamson and Yuxuan Wang and DeLiang Wang},
  title        = {Complex Ratio Masking for Monaural Speech Separation},
  journal      = {IEEE/ACM Transactions on Audio, Speech, and Language
                  Processing},
  volume       = {24},
  number       = {3},
  pages        = {483--492},
  year         = {2016},
  doi          = {10.1109/TASLP.2015.2512042}
}

@article{axialattention,
  author       = {Jonathan Ho and Nal Kalchbrenner and Dirk Weissenborn and
                  Tim Salimans},
  title        = {Axial Attention in Multidimensional Transformers},
  journal      = {arXiv preprint arXiv:1912.12180},
  year         = {2019},
  doi          = {10.48550/arXiv.1912.12180}
}

@inproceedings{resnet,
  author       = {Kaiming He and Xiangyu Zhang and Shaoqing Ren and Jian Sun},
  title        = {Deep Residual Learning for Image Recognition},
  booktitle    = {IEEE Conference on Computer Vision and Pattern
                  Recognition},
  pages        = {770--778},
  year         = {2016},
  doi          = {10.1109/CVPR.2016.90},
  url          = {https://arxiv.org/abs/1512.03385}
}

@inproceedings{jiang2024mdctcodec,
  title={MDCTCodec: A lightweight MDCT-based neural audio codec towards high sampling rate and low bitrate scenarios},
  author={Jiang, Xiao-Hang and Ai, Yang and Zheng, Rui-Chen and Du, Hui-Peng and Lu, Ye-Xin and Ling, Zhen-Hua},
  booktitle={2024 IEEE Spoken Language Technology Workshop (SLT)},
  pages={540--547},
  year={2024},
  organization={IEEE}
}

@inproceedings{adamw,
  author       = {Ilya Loshchilov and Frank Hutter},
  title        = {Decoupled Weight Decay Regularization},
  booktitle    = {International Conference on Learning Representations},
  year         = {2019},
  url          = {https://arxiv.org/abs/1711.05101}
}

@article{mdx2021,
  author       = {Yuki Mitsufuji and Giorgio Fabbro and Stefan Uhlich and
                  Fabian-Robert St{\"o}ter and Alexandre D{\'e}fossez and
                  Minseok Kim and Woosung Choi and Chin-Yun Yu and Kin-Wai
                  Cheuk},
  title        = {Music Demixing Challenge 2021},
  journal      = {Frontiers in Signal Processing},
  volume       = {1},
  pages        = {808395},
  year         = {2022},
  doi          = {10.3389/frsip.2021.808395}
}

@article{sdx2023,
  author       = {Giorgio Fabbro and Stefan Uhlich and Chieh-Hsin Lai and
                  Woosung Choi and Marco Mart{\'i}nez-Ram{\'i}rez and
                  Weihsiang Liao and Igor Gadelha and Geraldo Ramos and Eddie
                  Hsu and Hugo Rodrigues and Fabian-Robert St{\"o}ter and
                  Alexandre D{\'e}fossez and Yi Luo and Jianwei Yu and Dipam
                  Chakraborty and Sharada Mohanty and Roman Solovyev and
                  Alexander Stempkovskiy and Tatiana Habruseva and Nabarun
                  Goswami and Tatsuya Harada and Minseok Kim and Jun Hyung Lee
                  and Yuanliang Dong and Xinran Zhang and Jiafeng Liu and Yuki
                  Mitsufuji},
  title        = {The Sound Demixing Challenge 2023---Music Demixing Track},
  journal      = {Transactions of the International Society for Music
                  Information Retrieval},
  volume       = {7},
  number       = {1},
  pages        = {63--84},
  year         = {2024},
  doi          = {10.5334/tismir.171}
}

@article{scipy2020,
  title   = {SciPy 1.0: Fundamental Algorithms for Scientific Computing in Python},
  author  = {Virtanen, Pauli and Gommers, Ralf and Oliphant, Travis E. and
             Haberland, Matt and Reddy, Tyler and Cournapeau, David and
             Burovski, Evgeni and Peterson, Pearu and Weckesser, Warren and
             Bright, Jonathan and van der Walt, St{\'e}fan J. and Brett, Matthew
             and Wilson, Joshua and Jarrod Millman, K. and Mayorov, Nikolay and
             Nelson, Andrew R. J. and Jones, Eric and Kern, Robert and
             Larson, Eric and Carey, CJ and Polat, {\c{C}}ag{\c{a}}n and
             Feng, Yu and Moore, Eric W. and VanderPlas, Jake and Laxalde, Denis
             and Perktold, Josef and Cimrman, Robert and Henriksen, Ian and
             Quintero, Eric A. and Harris, Charles R. and Archibald, Anne M.
             and Ribeiro, Antonio H. and Pedregosa, Fabian and van Mulbregt, Max},
  journal = {Nature Methods},
  volume  = {17},
  pages   = {261--272},
  year    = {2020},
  doi     = {10.1038/s41592-019-0686-2}
}

\end{document}